\documentclass[a4paper,11pt]{article}
\pdfoutput=1 

\usepackage{jcappub} 
\usepackage{amsfonts}
\usepackage[utf8]{inputenc}
\usepackage{graphicx}
\usepackage{dcolumn}
\usepackage{bm}
\usepackage{amsmath}
\usepackage{lipsum}
\usepackage{xcolor}
\usepackage{calc}
\usepackage{accents}
\usepackage{float}
\usepackage{pifont}
\usepackage{makecell}
\usepackage{caption}
\usepackage{subcaption}
\usepackage{epsfig}
\usepackage{bbold}
\usepackage{amssymb}
\usepackage{amsbsy}
\usepackage{color}
\usepackage{slashed}
\usepackage{afterpage}
\usepackage{psfrag}
\usepackage[noabbrev]{cleveref}
\usepackage{dsfont}
\usepackage{enumerate}
\usepackage[shortlabels]{enumitem}
\usepackage[utf8]{inputenc}
\usepackage{amsmath}
\usepackage{graphicx}
\usepackage{graphbox}
\usepackage{braket}
\usepackage{soul}
\usepackage{multirow}
\usepackage{hhline}
\usepackage{abbr}

\newcommand{\reffig}[1]{figure~\ref{#1}}
\newcommand{\reftab}[1]{table~\ref{#1}}
\newcommand{\refsec}[1]{section~\ref{#1}}
\newcommand{\refequ}[1]{eq.~(\ref{#1})}
\newcommand{\refapp}[1]{appendix~\ref{#1}}

\newcommand{\cmFAST}{{\tt 21cmFAST}}%
\newcommand{\DarkHistory}{{\tt DarkHistory}}%
\newcommand{\cmfish}{{\tt 21cmfish}}%
\newcommand{\cmCAST}{{\tt 21cmCAST}}%
\newcommand{\popII}{PopII}%
\newcommand{\popIII}{PopIII}%
\subheader{ULB-TH/23-09}
\title{\boldmath 21cm signal sensitivity to dark matter decay}

\author[a]{G. Facchinetti,}
\author[a,b]{L. Lopez-Honorez,}
\author[c, d]{ Y. Qin,}
\author[e]{and A. Mesinger}

\affiliation[a]{Service de Physique Th\'eorique, C.P. 225, Universit\'e Libre de Bruxelles,\\ Boulevard du Triomphe, B-1050 Brussels, Belgium}
\affiliation[b]{Theoretische Natuurkunde \& The International Solvay Institutes, \\ Vrije Universiteit Brussel, Pleinlaan 2, B-1050 Brussels, Belgium}
\affiliation[c]{School of Physics, University of Melbourne, Parkville, VIC 3010, Australia}
\affiliation[d]{ARC Centre of Excellence for All Sky Astrophysics in 3 Dimensions (ASTRO 3D)}
\affiliation[e]{Scuola Normale Superiore, 56126 Pisa, PI, Italy}

\emailAdd{gaetan.facchinetti@ulb.be}
\emailAdd{laura.lopez.honorez@ulb.be}
\emailAdd{yuxiang.l.qin@gmail.com}
\emailAdd{andrei.mesinger@sns.it}

\abstract{
The redshifted 21cm signal from the Cosmic Dawn is expected to provide unprecedented insights into early Universe astrophysics and cosmology. Here we explore how dark matter can heat the intergalactic medium before the first galaxies, leaving a distinctive imprint in the 21cm power spectrum. We provide the first dedicated Fisher matrix forecasts on the sensitivity of the Hydrogen Epoch of Reionization Array (HERA) telescope to dark matter decays. We show that with 1000 hours of observation, HERA has the potential to improve current cosmological constraints on the dark matter decay lifetime by up to three orders of magnitude. Even in extreme scenarios with strong X-ray emission from early-forming, metal-free galaxies, the bounds on the decay lifetime would be improved by up to two orders of magnitude.  Overall, HERA shall improve on existing limits for dark matter masses below $2$ GeV$/c^2$ for decays into $e^+e^-$ and below few MeV$/c^2$ for decays into photons.
}

\keywords{dark matter theory, particle physics - cosmology connection, physics of the early universe, cosmology of theories beyond the SM}
\arxivnumber{2308.16656}

\begin{document}
\maketitle
\flushbottom

\section{Introduction}
\label{sec:introduction}

Dark matter (DM) is a crucial component of our Universe that shapes its evolution. Cosmological probes are amongst the  powerful tools at our disposal to shed light on the nature of DM. In particular, the Cosmic Microwave Background (CMB)~\cite{Planck:2018vyg} currently provides  the strongest constraints on DM properties. From CMB data,  the DM relic density is determined at the percent level. The latter surpasses the baryonic matter density by more than a factor of five. 
Cosmological observations can also efficiently probe weak couplings between DM and standard model (SM) particles that would leave specific imprints. In that regards, they are complementary to astro-particle experiments~\cite{Planck:2018vyg,Liu:2020wqz}.

The CMB is especially useful at constraining DM annihilation or decay into SM particles, see e.g.,~\cite{Shull:1985, Adams:1998nr, Chen:2003gz,Padmanabhan:2005es,Slatyer:2009yq,Slatyer:2015kla, Lopez-Honorez:2013cua,Diamanti:2013bia,Poulin:2015pna, Bolliet:2020ofj,Capozzi:2023xie,Liu:2023nct}. Both scenarios result in an exotic injection of energy into the intergalactic medium (IGM). For most of the SM final states, this exotic injected energy partially ionizes hydrogen and helium, increasing the residual free electron fraction post recombination.
This has an observational imprint in the temperature and polarization anisotropy power spectra of the CMB\footnote{CMB spectral distortions can also probe DM energy injection through heating before recombination takes place, see e.g.~\cite{Lucca:2019rxf}.}, constraining DM decay lifetimes up to $\tau\sim 10^{-24}$~s~\cite{Slatyer:2016qyl} and dark matter annihilation efficiencies up to $p_{\rm ann}\sim 3 \times 10^{28} $ cm$^3$/s/GeV~\cite{Planck:2018vyg}.\footnote{The parameter $p_{\rm ann}$ is defined as $ p_{\rm ann}=f_{\rm   eff}\langle \sigma v\rangle/m_{\rm DM}$ where $\langle \sigma v\rangle$  is the DM s-wave annihilation cross-section, $m_{\rm DM}$ is the DM mass and $f_{\rm eff}$ is the fraction of the energy released by the annihilation process that is effectively transferred to ionization around the redshifts to which the CMB anisotropy data are most sensitive.}

Late time probes (sensitive at $z\ll 1000$), including the Lyman-$\alpha$ forest in Quasi-Stellar Objects (QSO) spectra and the 21cm signal of the hyperfine HI transition, are expected to be particularly efficient in testing late time energy injection. They are more sensitive to the IGM temperature $T_{\rm k}$ and, consequently, to exotic energy injections transferred to the IGM in the form of heating~\cite{Valdes:2009cq,Evoli:2014pva,Lopez-Honorez:2016sur,Liu:2016cnk, Liu:2018uzy, DAmico:2018sxd,Cheung:2018vww, Mitridate:2018iag,Clark:2018ghm,Bolliet:2020ofj,Acharya:2023ygd,Liu:2023nct, Hiroshima:2021bxn, Mittal:2021egv}. In particular, the authors of Ref.~\cite{Liu:2020wqz} have used the Lyman-$\alpha$ forest sensitivity to $T_{\rm k}$ at redshifts $z\sim 4-6$ to derive constraints on decaying DM, see also~\cite{Cirelli:2009bb,Diamanti:2013bia,Liu:2016cnk,Liu:2020wqz}. They have improved on the CMB bounds for decaying DM lifetimes, disfavoring to $\tau\sim 10^{-25}$~s for DM decays into electron-positron pairs for $m_{\rm DM}<$ MeV$/c^2$. On the other hand, the 21cm signal will be sensitive to $T_{\rm k}$ at even earlier times, during the so-called Cosmic Dawn (CD) of galaxies ($z\sim$ 10-20) and the epoch of reionisation (EoR, $z\sim$ 5-10). Interestingly, the relative dearth of galaxies during the CD should make it easier to isolate an additional heating contribution from DM decay or annihilation. \\

Here we revisit the imprint of DM decays on the cosmological 21cm signal and obtain  the first forecasts of DM lifetime constraints that will be enabled by 21cm power spectrum measurements.
In particular, we focus on the Hydrogen Epoch of Reionisation Array (HERA) telescope which was designed to measure the 21cm power spectrum at a high signal to noise ratio (S/N).  HERA has completed deployment \cite{HERA:2021bsv} and is currently analysing data from an extended observational campaign.  An initial observational result performed with $71$ antennas (out of the total 331) and only 94 nights of measurement has already provided the most constraining upper bounds on the 21 cm power spectrum at redshifts $z=8$ and 10~\cite{HERA:2022wmy}. Combined with complementary observations of galaxy UV luminosity functions and the timing of the EoR, these upper limits imply significant early IGM heating \citep{HERA:2022wmy}. If this heating was provided by high mass X-ray binary stars in CD galaxies, expected to dominate the X-ray background at high redshifts (e.g., \cite{Fragos2013ApJ...776L..31F}), they would need to be considerably brighter than those observed today (perhaps because they were born in extremely metal poor environments; see e.g., \cite{Kaur2022MNRAS.513.5097K}). This  clearly illustrates the potential of 21-cm measurements to constrain IGM heating during the Cosmic Dawn. Consequently, it also indicates that 21cm cosmology will soon be mature enough to quantitatively probe exotic heating scenarios like decaying DM models. \\

Our numerical approach is built on the {\tt 21cmFAST} \cite{Mesinger:2010ne, Murray:2020trn} and the {\tt DarkHistory} packages \cite{Liu:2019bbm, Sun:2022djj}, which we combine into a hybrid code, {\tt exo21cmFAST}\footnote{The resulting code is available at \url{https://github.com/gaetanfacchinetti/exo21cmFAST} and should be merged with the main \cmFAST{} branch in a future release.}. We organize this paper as follows. In \refsec{sec:21cm_physics}, we briefly review 21cm cosmology and how to model and simulate the 21cm signal with {\tt exo21cmFAST}. In \refsec{sec:DM_decay}, we show the imprint of DM decay on the IGM and the corresponding 21cm signal. In \refsec{sec:method}, we present Fisher forecasts of constraints available with a 1000h observation using HERA.  We forecast joint constraints on decaying DM properties and two different models of CD galaxies.  Finally, in \refsec{sec:conclusion}, we discuss these results and conclude. In this work, we use a $\Lambda$CDM cosmology with parameters ($\Omega_{\mathrm{m}}, \Omega_{\mathrm{b}}, \Omega_{\mathrm{\Lambda}}, h, \sigma_8$, $n_s$) = (0.31, 0.049, 0.69, 0.68, 0.81, 0.97) following the {\it TT,TE,EE+lowE+lensing+BAO} result of Planck 2018 \cite{Planck2020A&A...641A...6P}). 

\section{A glimpse into 21cm cosmology}
\label{sec:21cm_physics}

We start with a brief review of 21-cm cosmology in \refsec{sec:21cm_physics:signal} before outlining in \refsec{sec:21cmbasic} our model for the astrophysics of first galaxies, which we then expand to include imprints of DM decays.

\subsection{The 21cm signal and its power spectrum}
\label{sec:21cm_physics:signal}

The redshifted  cosmic 21cm signal, arising from the hyperfine spin-flip transition of neutral hydrogen, can be seen in emission or in absorption compared  to the radio background. Here we fix the latter to the CMB whose temperature is denoted by $T_\mathrm{CMB}(z)$. The differential   brightness temperature of the 21cm signal can be expressed as \cite{Furlanetto2006PhR...433..181F}
\begin{equation}
		\delta T_\mathrm{b} \approx 20\mathrm{mK} \left(1-\frac{\mathrm{T_\mathrm{CMB}}}{T_\mathrm{S}}\right) x_{\mathrm{HI}} (1+\delta_{\rm b}) \left(1+\frac{1}{H}\,\frac{\mathrm{d}v_r}{\mathrm{d}r}\right)^{-1}\sqrt{\frac{1+z}{10}\frac{0.15}{\Omega_\mathrm{m}h^2}}\frac{\Omega_\mathrm{b}h^2}{0.023},
 \label{eq:deltaTb}
\end{equation}
where $T_{\rm S}$ is the spin temperature of neutral hydrogen in the IGM, $x_{\rm HI}$ is the neutral fraction, $\delta_b$ is the relative density perturbation in the baryon number density, $\Omega_{\rm m,b}$ are the matter and baryon energy densities relative to the critical energy density today and $h$ is the Hubble parameter in units of 100 km/s/Mpc. The relative motion of the neutral gas with regard to the Hubble flow, with $H$ denoting the Hubble rate, is taken into account with the $\frac{\mathrm{d}v_r}{\mathrm{d}r}$ term that is the gradient of the proper velocity along the line of
sight.\footnote{We note that equation \ref{eq:deltaTb} and the corresponding redshift space distortion term come from a first order Taylor expansion, in the limit of a very small 21cm optical depth.  Equation \ref{eq:deltaTb} is useful for building intuition, but we note that the {\tt 21cmFAST} code computes the full optical depth as well as includes redshift space distortions via a non-linear, sub-grid scheme (e.g., \cite{Greig:2015qca,Mao2012MNRAS.422..926M,Jensen2013MNRAS.435..460J}).} The spin temperature, $T_{\rm S}$, quantifies the relative occupancy of the two hyperfine levels of the ground state of neutral hydrogen. It is obtained from the equilibrium balance  of (i) absorption/emission of 21cm photons from/to the CMB background at a temperature $T_\mathrm{CMB}$;  (ii) collisions with atoms and electrons in the IGM, with the gas kinetic temperature denoted by $T_{\rm k}$; and (iii) resonant scattering of Lyman-$\alpha$ photons coupling the spin temperature to $T_{\rm k}$. When the spin temperature is coupled to IGM gas kinetic temperature through (ii) or (iii),  $T_{\rm S}$ can differ from  $T_\mathrm{CMB}$. The signal can appear in absorption (if $T_{\rm S}<T_{\rm CMB}$), in emission (if $T_{\rm S}>T_{\rm CMB}$) or can be zero (if $T_{\rm S}=T_{\rm CMB}$ and/or $x_{\rm HI}=0$). 

Spatial variation of IGM properties leads to fluctuations in the 21cm signal. In what follows, we refer to the 21cm global signal, $\overline{ \delta T_b}$, as the sky averaged brightness temperature while the 21cm power spectrum refers to the dimensional quantity, $\overline{\delta T_b^2} \Delta_{21}^2$, obtained from:
\begin{equation}
    \overline{\delta T_b^2} \Delta_{21}^2(k,z)=\overline{\delta T_b^2(z)} \times \frac{k^3}{2\pi^2} P_{21}(k,z)
    \label{eq:PS}
\end{equation}
where  $P_{21}$ is defined as:
\begin{equation}
    \langle  \tilde\delta_{21} ({\bf k},z) \tilde\delta_{21} ({\bf k'},z) \rangle= (2\pi)^3 \delta^D({\bf k}- {\bf k'}) P_{21}(k,z)
\end{equation}
with $\langle \rangle$ the ensemble average, $\bf k$ the comoving wave vector, and $\tilde\delta_{21} ({\bf k},z)$  the Fourier transform of $\delta_{21} ({\bf x},z)= {\delta T_b}({\bf x},z)/\overline{\delta T_b}(z) - 1$. Notice that $\bf x$ denotes the position vector. 

\subsection{Modelling the 21cm signal}
\label{sec:21cmbasic}

As seen above, the 21cm signal depends on the IGM temperature and ionization fraction. These quantities, in turn, are influenced by the radiation emitted by stars and sources of exotic energy in the late Universe. Here we review how astrophysical sources are modeled in {\tt 21cmFAST}, before moving onto the inclusion of exotic heating in our modified version, {\tt exo21cmFAST}, in the following section. 

\subsubsection{Ionization, excitation and heating of the IGM}

The EoR is an inhomogeneous process, with ionizing photons from galaxies carving out cosmic HII regions that grow and eventually overlap.  We denote the volume filling factor of HII regions as $Q_{\rm HII}(z)$ (thus filling factor of the mostly neutral regions is $1-Q_{\rm HII}$).
The {\tt 21cmFAST} codes employ an excursion-set algorithm\footnote{The excursion-set algorithm evaluates the \textit{average} photon budget within spherical regions. In this work, ${n}_{\rm ion}$, ${n}_{\rm rec}$, ${x}_e$ and etc are smoothed quantities over the considered areas (see more in \cite{Mesinger:2010ne,Qin2020MNRAS.495..123Q}).}  to compute inhomogeneous reionization \cite{Furlanetto2004ApJ...613....1F}.
 A cell is considered as ionized when
\begin{equation}
   {n}_{\rm ion} \ge (1+{n}_{\rm rec})(1-{x}_e)\,,
\label{eq:nion}
\end{equation}
where  ${n}_{\rm ion}$  denotes the cumulative number of ionizing photons per baryon (see \refsec{sec:starform}), ${n}_{\rm rec}$ gives the cumulative number of recombinations per baryon  within spheres of decreasing radii and ${x}_e$ accounts for secondary ionization in mostly-neutral IGM. The redshift evolution of latter quantity and of  the gas kinetic temperature
($T_{\rm k}$) is described by
 \begin{eqnarray}
 \frac{dx_e }{dz} &=& \frac{dt}{dz} \left(\Lambda_{\rm ion}^{\rm X} + \Lambda_{\rm ion}^{\rm DM}
 - \alpha_{\rm A} \, C \, x_e^2 \, n_{\rm b} \, \mathfrak{f}_{\rm H} \right)~, 
 \label{eq:xe} \\
 \frac{d T_{\rm k}}{dz} & = &\frac{2}{3 \, k_{\rm B} \, (1+x_e)} \, \frac{dt}{dz} \, \sum_\beta \epsilon^\beta_{\rm heat} 
 + \frac{2 \, T_{\rm k}}{3 \, n_{\rm b}} \, \frac{dn_{\rm b}}{dz}  - \frac{T_{\rm k}}{1+x_e} \, \frac{dx_e}{dz} ~,
 \label{eq:TK}
\end{eqnarray}
where $k_{\rm B}$ is the Boltzmann constant, $\epsilon^\beta_{\rm heat}({\bf x}, z)$ is the heating rate per baryon from different sources $\beta$ including Compton scattering (effective at high redshifts, $z \gtrsim 300$) as well as heating by X-rays ($\epsilon_{\rm heat}^{\rm X}$) and DM ($\epsilon_{\rm heat}^{\rm DM}$). $\Lambda_{\rm ion}^{\rm X,DM}$ are the secondary ionization rates per baryon accounting for the contribution from either X-rays or DM, $\alpha_{\rm A}$ is the case-A recombination coefficient, $C\equiv \langle n_e^2 \rangle / \langle n_e \rangle^2$ is the clumping factor, set to its default value $C=2$ in our analysis, with $n_e$ the electron number density, and $\mathfrak{f}_{\rm H}=n_{\rm H}/n_{\rm b}$ is the hydrogen number fraction. The implementation of $\Lambda_{\rm ion}^{\rm DM}$ and $\epsilon_{\rm heat}^{\rm DM}$ accounting for DM energy injection is specific to {\tt exo21cmFAST} which is detailed in section \ref{sec:DM_decay}. When considering ionization in both the mostly-ionized and mostly-neutral IGM, the total mean ionization fraction becomes:
\begin{equation}
    x_i \approx Q_{\rm HII} + x_e (1-Q_{\rm HII}).
   \label{xtot}
\end{equation}
The strength of the Wouthuysen-Field effect that couples the spin temperature of gas to its kinetic temperature is determined by the total Ly$\alpha$ background, which has the following components:
\begin{equation}
J_{\alpha} = J_{\alpha}^{X} + J_{\alpha}^{\star} + J_{\alpha}^{\rm DM} ~,
\label{eq:Jalpha}
\end{equation}
where $J_{\alpha}^X$ arises from X-ray sources,  $J_{\alpha}^\star$ encapsulates the contribution from stellar photons with energies between Lyman-$\alpha$ and the Lyman limit, while $J_{\alpha}^{\rm DM}$ accounts for the DM contribution introduced in \refsec{sec:DM_decay}. We then evaluate the Lyman-$\alpha$ coupling efficiency according to
\begin{equation}
    x_\alpha = \frac{1.7\times10^{11}}{1+z} \left( \frac{J_\alpha}{{\rm s^{-1}Hz^{-1}cm^{-2}sr^{-1}}} \right) S_\alpha,
\end{equation}
where $S_\alpha$ acts as a quantum mechanical correction of the order of unity \cite{Hirata2006MNRAS.367..259H} and further compute the spin temperature as
\begin{equation}
    T_{\rm S}^{-1} = \frac{T_{\rm CMB}^{-1} + (x_\alpha + x_c)T_{\rm k}^{-1}}{1+x_\alpha + x_c},
\end{equation}
with $x_c$ denoting the spatially-varying collisional coupling efficiency in the IGM.

In the next subsections, we discuss our parameterization of galaxy properties, which allows us to determine the budget of photons in various wavelength ranges and evaluate the aforementioned radiative backgrounds that are essential to the 21cm signal. 

\subsubsection{Star formation and galaxy evolution}
\label{sec:starform}

As we wish to consider the 21cm signal during the initial stages of the CD, our models will include the very first galaxies, hosted by so-called minihalos (with a virial mass $M_{\rm vir} \lesssim 10^8{\rm M}_\odot$) for which the dominant cooling channel is provided by roto-vibrational transitions of H$_2$.  Due to their metal free gas and shallower potential wells, these molecular-cooling galaxies (MCGs) might have different properties compared to the atomic-cooling galaxies (ACGs) that we have observed at later times. They are expected to predominately host Population-III stars\footnote{In this work, we consider ACGs/MCGs to host mostly PopII/PopIII stars and use superscripts II/III to distinguish the two populations.} and likely possess different initial mass functions (IMFs), as well as potentially different star formation efficiencies (see e.g., \cite{Brommdoi:10.1146/annurev.astro.42.053102.134034,Xu2016ApJ...823..140X,Schneider_2002,Mebane10.1093/mnras/sty1833} and references therein). 

The star formation rate (SFR) of an ACG/MCG inside a halo of virial mass $M_{\rm vir}$ is parameterised by \cite{Qin2020MNRAS.495..123Q}
\begin{equation}
    \dot M_\star^{\rm II/III} \equiv \frac{f_\star^{\rm II/III}   M_{\rm vir} }{t_\star H(z)^{-1}}
\end{equation}
where $f_\star^{\rm II}$ and $f_\star^{\rm III}$ represent different stellar-to-halo mass ratios for ACGs and MCGs, respectively. Here, $t_{\star} H^{-1}$ is a characteristic star-formation time-scale, with $t_\star$ a dimensionless parameter taking values between zero and one (for our analysis we chose a fiducial value $t_\star = 0.5$, see \reftab{tab:params} for all fiducial values). Through $t_\star$, the model encompasses scenarios where all stars are formed in an instantaneous burst event or as a gradual buildup over the age of the Universe. References~\cite{Stefanon2021,Shuntov2022A&A...664A..61S} have further inferred a nearly non-evolving, power-law relation between $f_\star^{\rm II/III}$ and $M_{\rm vir}$ for galaxies that are expected to dominate reionization. Therefore, we consider
\begin{equation} 
    \begin{split}
    & f_\star^{\rm II} = \frac{\Omega_{\rm b}}{\Omega_{\rm m}} \min\left\{1, f_{\star,10}^{\rm II} \left(\frac{M_{\rm vir}}{10^{10}~{\rm M}_\odot}\right)^{\alpha_\star^{\rm II}}\right\}  \\
    & f_\star^{\rm III} = \frac{\Omega_{\rm b}}{\Omega_{\rm m}} \min\left\{1, f_{\star,7}^{\rm III} \left(\frac{M_{\rm vir}}{10^{7}~{\rm M}_\odot}\right)^{\alpha_\star^{\rm III}}\right\}\,,
    \end{split}
    \label{eq:fstar}
\end{equation}
where $f_{\star,10}^{\rm II}$, $f_{\star,7}^{\rm III}$, $\alpha_{\star}^{\rm II}$ and $\alpha_{\star}^{\rm III}$ are free parameters. Similarly, the UV ionizing escape fractions of ACGs and MCGs are also parameterized as power-law relations with the halo mass
\begin{equation}
    \begin{split}
    & f_{\rm esc}^{\rm II} =  \min\left\{1, f_{\rm esc,10}^{\rm II} \left(\frac{M_{\rm vir}}{10^{10}~{\rm M}_\odot}\right)^{\alpha_{\rm esc}}\right\} \\
    & f_{\rm esc}^{\rm III} =  \min\left\{1, f_{\rm esc,7}^{\rm III} \left(\frac{M_{\rm vir}}{10^{7}~{\rm M}_\odot}\right)^{\alpha_{\rm esc}}\right\}\, 
    \end{split}
    \label{eq:fesc}
\end{equation}
where $f_{\rm esc,10}^{\rm II}$, $f_{\rm esc,7}^{\rm III}$ and $\alpha_{\rm esc}$ are additional free parameters (for simplicity, the same scaling index is considered in this work for $f_{\rm esc}^{\rm II/III}$). The efficiency parameters, $f_{\star,10}^{\rm II}$, $f_{\star,7}^{\rm III}$, $f_{\rm esc,10}^{\rm II}$, $f_{\rm esc,7}^{\rm III}$ can vary between zero and one.\footnote{In practice (e.g., for a Bayesian inference), because of the broad range of associated uncertainties, $f_{\star,10}^{\rm II}$, $f_{\star,7}^{\rm III}$, $f_{\rm esc,10}^{\rm II}$, $f_{\rm esc,7}^{\rm III}$ priors are often implemented in logarithmic space with a flat prior between e.g., $10^{-3}$ and $10^0$.} In our Fisher Matrix analysis, we set $\log_{10} f_{\star,10}^{\rm II}=-1.3$ and $\log_{10} f_{{\rm esc},10}^{\rm II}=-1.0$ while we take $\log_{10} f_{\star,7}^{\rm III}=-2.0$ and $\log_{10} f_{{\rm esc},7}^{\rm III}=-2.0$ following~\cite{Park:2018ljd} and \cite{Qin2020MNRAS.495..123Q}.  On the other hand, values of the scaling index parameters are mostly guided by observational results or detailed simulations. For instance, fitting their modelled stellar mass function to observations, the authors of Ref.~\cite{Mutch2016MNRAS.462..250M,Park:2018ljd} find that $\alpha_\star^{\rm II}$ is close to 0.5. This result is consistent with a supernova-regulated galaxy growth history \cite{Wyithe2013MNRAS.428.2741W}. In addition, $\alpha_{\rm esc}$ is likely negative as we expect more low-column density channels to be created by supernovae in low-mass galaxies and to allow UV ionizing photons to escape into the IGM. In our Fisher Matrix analysis, we set $\alpha_{\star}^{\rm II}= \alpha_{\star}^{\rm III}=0.5$ and $\alpha_{\rm esc}=-0.5$.

The number density of ACGs and MCGs per unit halo mass can be described by the halo mass function (HMF)\footnote{In this work, the Sheth, Mo and Tormen (SMT) HMF \cite{SMT2001MNRAS.323....1S} is adopted.} (${{\rm d}n}/{{\rm d}M_{\rm vir}}$) weighted by different duty cycles ($f_{\rm duty}^{\rm II/III}$). The duty cycles account for inefficient star formation in low mass ACGs and MCGs. This mass threshold for ACGs is  given by either the atomic cooling limit ($M_{\rm atom}$) or by a characteristic mass ($M_{\rm crit}^{\rm RE}$) below which photoheating during reionization is able to significantly suppress gas content \citep{Sobacchi2013MNRAS.432.3340S}. The duty cycle of ACGs can be described by~\cite{Park:2018ljd,Qin2020MNRAS.495..123Q} 
\begin{equation}
     f_{\rm duty}^{\rm II}\equiv\exp\left(-\frac{M_{\rm turn}^{\rm II}}{M_{\rm vir}}\right).
     \label{eq:fduty}
 \end{equation} 
 When considering both ACGs and MCGs in our analysis, we self-consistently evaluate $ M_{\rm turn}^{\rm II}$ as $\max\left(M_{\rm atom}, M_{\rm crit}^{\rm RE}\right)$ where $M_{\rm atom}=5\times 10^7 {\rm M}_\odot ((1+z)/10)^{-1.5}$ \cite{BARKANA2001125}.
 Furthermore, the duty cycle of MCGs can be written as~\cite{Qin2020MNRAS.495..123Q,Munoz2022MNRAS.511.3657M} 
\begin{equation}
    f_{\rm duty}^{\rm III} \equiv \exp\left(-\frac{M_{\rm turn}^{\rm III}}{M_{\rm vir}}\right) \exp\left(-\frac{M_{\rm vir}}{M_{\rm atom}}\right)\,. 
    \label{eq:fduty_III}
\end{equation}
Compared to \refequ{eq:fduty}, the  additional factor of $\exp\left(-M_{\rm vir}/M_{\rm atom}\right)$ results from the assumption that there is a smooth
transition from MCGs (below $M_{\rm atom}$) hosting PopIII stars to
ACGs (above that mass) hosting PopII stars. The lower mass cutoff $M_{\rm turn}^{\rm III}$, for MCGs star formation, is set by three types of feedback mechanisms. Similarly, MCGs' gas reservoir also becomes inadequate when their host halos are too small (i.e., $<M_{\rm crit}^{\rm RE}$) to withstand the UV ionizing background. In addition to this, as molecular hydrogen can be easily dissociated by photons within the Lyman-Werner (LW) energy range ($11.2-13.6$ eV), star formation is also quenched in regions that have a LW background above a critical value $M_{\rm crit}^{\rm LW}$. In our analysis, we have thus $M_{\rm turn}^{\rm III}= \max\left(M_{\rm crit}^{\rm RE}, M_{\rm crit}^{\rm LW}\right)$ and evaluate these critical masses using the excursion algorithm~\cite{Sobacchi2013MNRAS.432.3340S}.\footnote{Upon the completion of this work, the authors of~\cite{Qin:2023kkk} have performed an initial study of the impact of DM energy injection on the $H_2$ content of first galaxies. They show that halo collapse and star formation can be affected, but the direction of the effect depend on the DM lifetime, redshift and astrophysics (LW self-shielding). A detailed study of these effects would require hydrodynamical simulations of $H_2$ formation with decay of DM. The impact of DM heating on the threshold for star formation is thus not included in our analysis.}

Note that in this work, we first consider a simplified model in which ACGs and MCGs follow the same scaling relations. In this case, it is effectively a single-population model and therefore $ f_{\star,10}^{\rm II}$, $\alpha_\star^{\rm II}$, $f_{\rm esc,10}^{\rm II}$, $\alpha_{\rm esc}$ are sufficient to describe their stellar mass/UV ionizing escape fractions. In addition, the threshold for star forming is set by  $f_{\rm duty}^{\rm II}$ as  in \refequ{eq:fduty} with $M_{\rm turn}^{\rm II}=M_{\rm turn}$ taken as a free parameter for more flexible interpretation of the feedback strength. In models without MCGs, $M_{\rm turn}$ is set above $10^8~{\rm M}_\odot$ due to the atomic cooling limit at these redshifts. In addition, currently observed galaxy UV luminosity functions \cite{Bouwens2015a,Bouwens2016,Oesch2018ApJ...855..105O} indicate $M_{\rm turn}$ is likely to be lower than $3\times10^9~{\rm M}_\odot$ \cite{Park:2018ljd}. In our Fisher Matrix analysis, we will use the fiducial value of  $M_{\rm turn}= 10^{8.7}~{\rm M}_\odot$. \\

With these definitions, we can evaluate the UV ionizing photon budget from  \refequ{eq:nion}:
\begin{equation}
    n_{\rm ion} = \frac{1}{\rho_b}\sum_{i\in{\rm \{II,III\}}}\int {\rm d}M_{\rm vir} \frac{{\rm d}n}{{\rm d}M_{\rm vir}} f_{\rm duty}^{i} M_{\rm vir} f_\star^{i} f_{\rm esc}^{i} n_{\gamma}^{i},
\end{equation}
where $\rho_b$ is the baryon density, $n_{\gamma}^{\rm II/III}$ refers to the number of UV ionizing photons emitted per stellar baryon and, in this work, is chosen to be 5000 and 50000 for ACGs and MCGs, respectively. We can further estimate photoheating feedback and evaluate $M_{\rm crit}^{\rm RE}$ following \cite{Sobacchi2013MNRAS.432.3340S}. Similar approaches are also adopted to calculate the Lyman-$\alpha$ background from stellar components (i.e., $J_{\alpha}^*$ in equation \ref{eq:Jalpha}) as well as the LW radiation and its feedback strength ($M_{\rm crit}^{\rm LW}$). We refer readers who are interested in these details to Ref.~\cite{Qin2020MNRAS.495..123Q}.

\subsection{X-ray heating and its impact on the 21cm signal}
\label{sec:21numerics}

\begin{figure}[t!]
\centering
\includegraphics[width=0.49\linewidth]{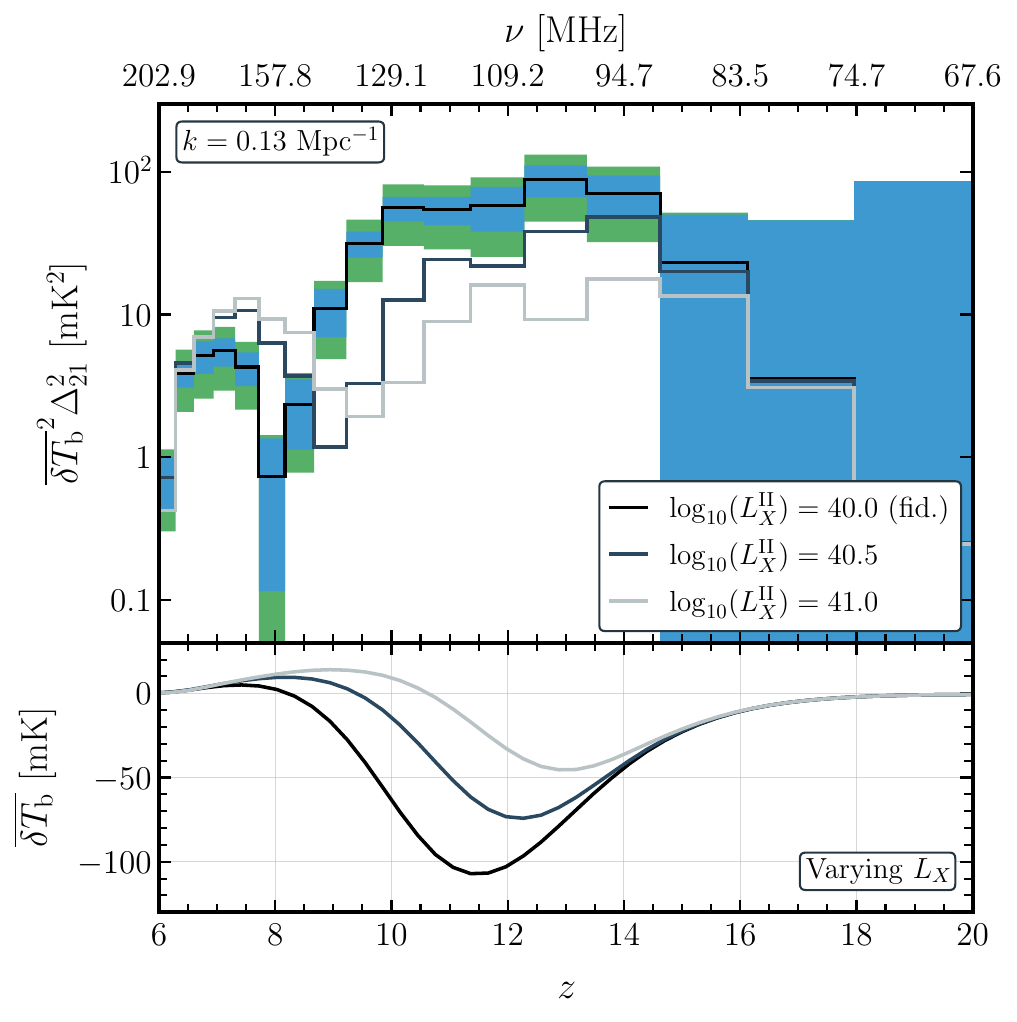}
\includegraphics[width=0.49\linewidth]{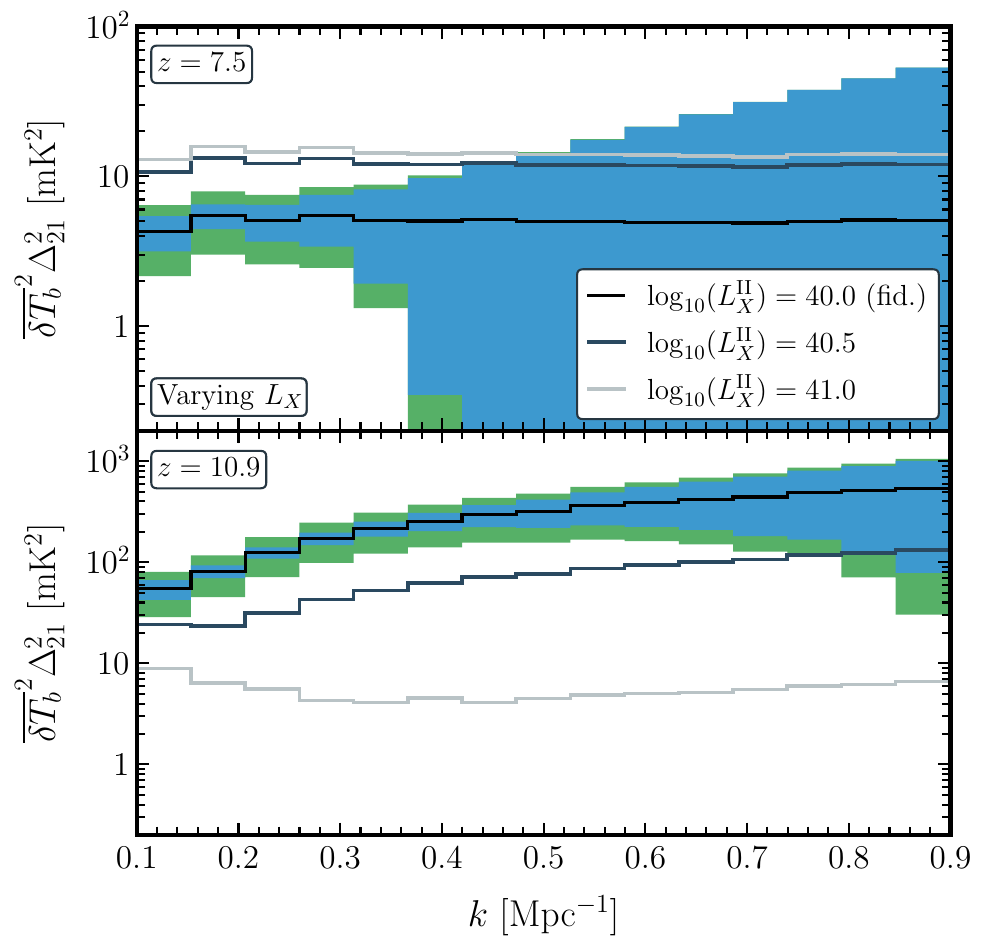}
\caption{21cm signal for 3 illustrative  scenarios. The fiducial model considered in this analysis is shown with a black line (with $\log_{10}  L_X^{\rm II}=40$) and 2 other scenarios with higher values of the X-ray luminosity amplitude are shown with gray lines (dark gray and light gray: $\log_{10}  L_X^{\rm II}=40.5$ and $41$ respectively). The blue region around the fiducial model corresponds to the 2-$\sigma$ thermal noise for a 1000h observation with HERA, while the green region additionally includes cosmic variance and modeling uncertainty (see \refsec{sec:method:Fisher_matrices} for details). {\bf Left panel:} 21cm power spectrum as a function of the redshift for a fixed scale, $k=0.13$/Mpc, relatively free of foregrounds (top), and the sky averaged differential brightness temperature (bottom). {\bf Right panel}: 21cm power spectrum as function of the scale at redshift $z=7.5$ (top) and $z=10.9$ (bottom). }
\label{fig:global-PS-LX}
\end{figure}

To estimate the heating due to X-rays as well as their contribution to ionizing the mostly-neutral IGM and coupling its spin temperature to the kinetic temperature, we further assume that the specific X-ray luminosity per unit star formation rate (denoted as ${\cal L}_X^{\rm II/III}$ in units of ${\rm s^{-1}}$ per ${\rm M_\odot\ yr^{-1}}$) follows power-law relations \cite{Das2017MNRAS.469.1166D} with an energy index of $\alpha_{\rm X}$, i.e., ${\cal L}_X^{\rm II/III} \propto E^{-\alpha_{\rm X}}$ (in this work $\alpha_{\rm X} = 1$). It is normalized with the integrated soft-band ($E<2$ keV) luminosity per SFR (in units of ${\rm erg\ yr\ s^{-1}\ M_\odot^{-1}}$), that we denote $L_{X}^{\rm II/III}$:
\begin{equation}
L_X^{\rm II/III} \equiv \int^ {2 {\rm keV}}_{E_0}
{\rm d} E \, {{\cal L}_X^{\rm II/III}}\,,
  \label{eq: LX}
\end{equation}
where $E_0$ is the X-ray energy threshold above which photons can not escape the host galaxy. $L_X^{\rm II}$ and $L_X^{\rm III}$ are considered as free parameters in our analysis while $L_X^{\rm III}$ is set equal to $L_X^{\rm II}$ in the single-population model. These allow us to evaluate the X-ray emissivity at the emitting redshift:
\begin{equation}
{\epsilon}_{X} =  \sum_{i\in\left\{{\rm II,III}\right\}}  \int {\rm d}M_{\rm vir} \frac{{\rm d}n}{{\rm d}M_{\rm vir}} f_{\rm duty}^{i} \dot{M}_\star^i  {\cal L}_X^{i}\,.
\label{eq:e_X}
\end{equation}

Note that, the X-ray threshold $E_0$ depends on the interstellar medium as well as the environment of X-ray sources (likely to be high mass X-ray binaries). Motivated by hydrodynamic simulations of the first galaxies~\cite{Das2017MNRAS.469.1166D}, $E_0$ is  assumed to range between 0.2 and 1.5 keV, which correspond to a column density of roughly $2\times10^{19}$ and $10^{23}~{\rm cm}^{-2}$, respectively. In our Fisher Matrix analysis, we take $E_0=0.5$ keV as a fiducial value.  On the other hand, $L_X^{\rm II/III}$ for high-redshift galaxies is less understood. Local, star-forming galaxies typically have $L_X{\sim}10^{39.5}$ \cite{Mineo2012MNRAS.419.2095M}, though  CD galaxies due to their lower metallicity ISM are expected to have higher values of $L_X$ (e.g., \cite{Lehmer2021ApJ...907...17L,Fragos2013ApJ...764...41F,Brorby10.1093/mnras/stw284}). We set $\log_{10}L_X^{\rm II/III}=40$ as the fiducial model in this work, motivated by recent analyses of the HERA observation \cite{HERA:2022wmy,Lazare2023arXiv230715577L}. The consequences of a larger value $\log_{10}L_X^{\rm II}=41$ are discussed in \refapp{app:impact_LX_forecast}. 

For illustration, \reffig{fig:global-PS-LX} shows the 21cm signal for three values of $L_X^{\rm II}$  considering a single population of galaxies. The black line corresponds to a fiducial value of $\log_{10}L_X^{\rm II}=40$.  The blue region around the fiducial model corresponds to the 2$\sigma$ thermal noise for a 1000h observation with HERA, while the green region additionally includes cosmic variance and modeling uncertainty (see \refsec{sec:method:Fisher_matrices} for details). In the left panel, we show the power spectrum (top) at a fixed scale $k=0.13~{\rm Mpc}^{-1}$ (that is expected to be relatively free from foreground contamination) and the global signal (bottom) as a function of redshift. 

Using the fiducial model as benchmark, we see that the global signal (bottom plot) displays an absorption trough between $z\sim 16$ and 9, and in emission below $z\sim 8$. Very roughly,  the minimum of absorption separates the so-called epoch of Lyman-$\alpha$ or Wouthuysen–Field (WF) coupling ($z\gtrsim 11$) and  the subsequent epoch of heating (EoH, for $z\lesssim 11$). Between $z\sim 8$ and $z\sim 6$, the ionized fraction of IGM increases, i.e., $x_{\rm HI}$ decreases,  correspoding to the EoR. Around $z\sim 6$ the IGM is fully ionized and $\overline{\delta T_b}$ becomes zero.  In parallel, we see that the power spectrum on these large scales, i.e.,  $k\lesssim 0.13/$Mpc,  (upper plot) displays multiple peaks as a function of  redshift.  As discussed in previous works (e.g., \cite{Lidz:2006vj, Pritchard:2011xb,Mesinger2014MNRAS.439.3262M}), three peaks in the evolution of the large-scale power  correspond  to the three astrophysical epochs: WF coupling, EoH and EoR, when the 21cm power spectrum is dominated by the auto-power spectrum of $x_\alpha$, $T_K$ and $Q_{\rm HII}$, respectively.  Between these epochs, the corresponding cross power spectra are negative and have a significant contribution, causing the large scale power to drop.\footnote{Note that to leading order, the power spectrum of the product of two fields (ionization, density field etc), $A$ and $B$, can be written as a sum of their auto and cross power: $P_{AA} + P_{BB} + 2P_{AB}$.} For example, the first regions to ionize are those in the proximity of galaxies that are also exposed to a stronger X-ray background and are hotter than average.  Thus, reionization effectively zeros the peaks of the 21cm signal during the EoH. This causes the large scale power to decrease, before increasing again when fluctuations in $Q_{\rm HII}$ begin to dominate during the middle stages of the EoR.
  
Increasing $\log_{10} L_X^{\rm II}$ shifts the EoH to earlier times.  This decreases the overlap between the EoR and EoH, diminishing the importance of the cross terms and thus increasing the peak power spectrum amplitude during the EoR.  However, it increases the overlap between the EoH and WF coupling, boosting cross terms and decreasing the associated peak power during the EoH.  Similar effects (increase of the EoR peak and suppression of power in EoH)  can be observed when considering two populations of galaxies and increasing $L_X^{\rm III}$, the normalisation of X-ray flux from MCGs. This is illustrated in \refapp{app:MCGs}.\footnote{Note that when decreasing $\log_{10} L_X^{\rm II}$ to values  much lower than ${\sim} 39$, the signal might never appear in emission before the EoR (e.g., \cite{Park:2018ljd}).  In this case the contrast between the ionized IGM (with zero signal) and the neutral IGM (with $\delta T_b$ strongly negative) becomes very large (so-called "cold reionization" \cite{Mesinger:2013nua} ).  This is the reason why HERA could set a lower bound on the X-ray luminosity with only preliminary data, see~\cite{HERA:2021noe, HERA:2022wmy}.}

\section{The impact of Dark matter decay on the 21cm signal}
\label{sec:DM_decay}

Dark matter annihilations and decays inject energy into the IGM. This induces extra heating, ionization and excitation of the IGM that leave a footprint in the 21cm signal. In this paper, we focus on dark matter decays for two reasons. First, there is a large room for improvement on current constraints as DM decay mainly impacts cosmological observables at relatively late times (at $z\ll 1000$) while DM annihilation is already be strongly constrained by CMB observations (see e.g., the discussion in ~\cite{Diamanti:2013bia,Liu:2016cnk,Planck:2018vyg,Planck2020A&A...641A...6P}). In addition, the DM annihilation signature in the 21cm signal is expected to strongly depend on the late time boost arising from structure formation (see e.g.,~\cite{Evoli:2014pva,Lopez-Honorez:2016sur}), contrary to CMB bounds \cite{Lopez-Honorez:2013cua, Poulin:2015pna}. We therefore leave the study of DM annihilation for future work.

\subsection{Dark matter energy injection and deposition}
\label{sec:DM_decay:energy_injection}

A priori, DM can decay into a plethora of SM final states. Yet, after subsequent decays, hadronization processes, etc., it is only the leftover photons, electrons, and positrons that will efficiently deposit their energy into the IGM (see e.g.,~\cite{Padmanabhan:2005es,Slatyer:2012yq,Slatyer:2015jla,Slatyer:2015kla}). For this reason, we will  focus on decays into electron-positron pairs and photon pairs.

High energy SM final state particles injected into the IGM from DM decays do not deposit their energy instantaneously. In addition, multiple channels (denoted with a subscript $a$) of energy deposition have to be considered. These channels are IGM heating (with $a=$ heat), Hydrogen ionization ($a=$ HII), Helium single or double ionization ($a=$ HeII or HeIII), and neutral atom excitation ($a=$ exc). The energy deposition rate (per average baryon number) can be expressed in terms of the energy injection rate as~\cite{Slatyer:2009yq}
\begin{equation}
	 \epsilon_a^{\rm DM} \equiv \frac{1}{n_{\rm b}(z)}\left(\frac{dE_a (x_e,z)}{dt\,dV}\right)_{\rm deposited} =   f_a(x_e,z) \frac{1}{n_{\rm b}(z)}\left(\frac{dE (z)}{dt\,dV}\right)_{\rm injected}\, ,
	\label{eq:dep}
\end{equation}
where $n_{\rm b}(z)$ denotes the baryon number density.  If all of the  dark matter  decays with a lifetime, $\tau$, longer than the age of the universe, the injected energy per unit time and volume takes the form:
 \begin{equation}
\left( \frac{d E}{d V d t}\right)_{\rm injected}=
   (1+z)^3 \, \frac{\rho_{\rm DM,0} c^2}{\tau }\,,
   \label{eq:dEdVdtinj}
\end{equation}
where $c$ denotes the speed of light. Below we also use the DM decay rate, $\Gamma\equiv 1/\tau$, to characterize DM energy injection.

The coefficients, $f_a(x_e, z)$, in \refequ{eq:dep} are the DM energy deposition efficiencies. They account for all of the details associated with the delay in energy deposition and separation into different channels $a$ at a given redshift $z$ and free-electron fraction $x_e$. In this analysis, we compute energy deposition efficiencies using the public {\tt DarkHistory} package~\cite{Liu:2019bbm}.\footnote{The version of DarkHistory used in this work does not account for absorption of soft photons with energies below 10.2 eV. In particular, heating from free-free interactions of such photons is neglected in our analysis, see e.g.~\cite{Acharya:2023ygd} accounting for this channel.} We have implemented, in {\tt exo21cmFAST}, a slightly adapted version of release 1.1 (with tables of the deposition fractions that have been upgraded from release 1.0) \cite{Sun:2022djj}. To account for early exotic energy injection into the IGM, \DarkHistory{} not only evaluates the deposition efficiencies but also solves for the homogeneous IGM evolution between redshift 3000 and 35. Afterwards, \cmFAST{} takes over to simulate the full IGM evolution including late-time astrophysical processes from cosmic dawn to reionisation. 

The energy deposition rate per channel per baryon $\epsilon_a^{\rm DM}$, defined in \refequ{eq:dep}, enters directly in the evolution of the ionized fraction, the IGM temperature and the Lyman-$\alpha$ flux \cite{Valdes:2009cq,Evoli:2014pva,Lopez-Honorez:2016sur}, which drive the 21cm signal (see \refsec{sec:21cm_physics}). 
Indeed, the energy injection from DM decays contributes to \refequ{eq:xe}, \refequ{eq:TK}, and \refequ{eq:Jalpha} via the heating rate  $\epsilon_{\rm heat}^{\rm DM}$, the ionization rate $\Lambda_{\rm ion}^{\rm DM}$, and the Ly$\alpha$ flux $J_\alpha^{\rm DM}$. While $\epsilon_{\rm heat}^{\rm DM}$ is directly computed from \refequ{eq:dep} by \DarkHistory{}, the DM-induced ionization rate and Ly$\alpha$ background flux are respectively derived as
\begin{eqnarray}
    \Lambda_{\rm ion}^{\rm DM} & = &  \mathfrak{f}_{\rm H} \, \frac{\epsilon^{\rm DM}_{\rm HII}}{E_{\rm th}^{\rm HI}} +\mathfrak{f}_{\rm He} \, \frac{\epsilon^{\rm DM}_{\rm HeII}}{E_{\rm th}^{\rm HeI}} \, ,\\
    J_{\alpha}^{\rm DM} & = &\frac{c\, n_{\rm b}}{4\pi H(z)\nu_\alpha} \frac{\epsilon_{\mathrm{Ly}\alpha }^{\rm DM}}{h\nu_\alpha}  ~.
    \label{eq:JalphDM}
\end{eqnarray}
\DarkHistory{} accounts for the variation of the deposition efficiencies with the ionization fraction, $x_e$. The DM source terms in \refequ{eq:xe} and \refequ{eq:TK}  are thus not constants but depend on the state of the IGM. The authors of \cite{Liu:2020wqz} refer to this as the \emph{backreaction} of the IGM on the deposition efficiencies. In \DarkHistory{} we can switch the backreaction off to gauge its significance by considering the IGM evolution that would result from a scenario without exotic energy injection, $\tilde x_e$ and forcing $f_a(x_e = \tilde x_e, z)$. For the models considered in our Fisher Matrix analysis we have noticed that  the backreaction has a negligible impact on the ionization and temperature evolution.  

\begin{figure}
\centering
\includegraphics[width=0.95\linewidth]{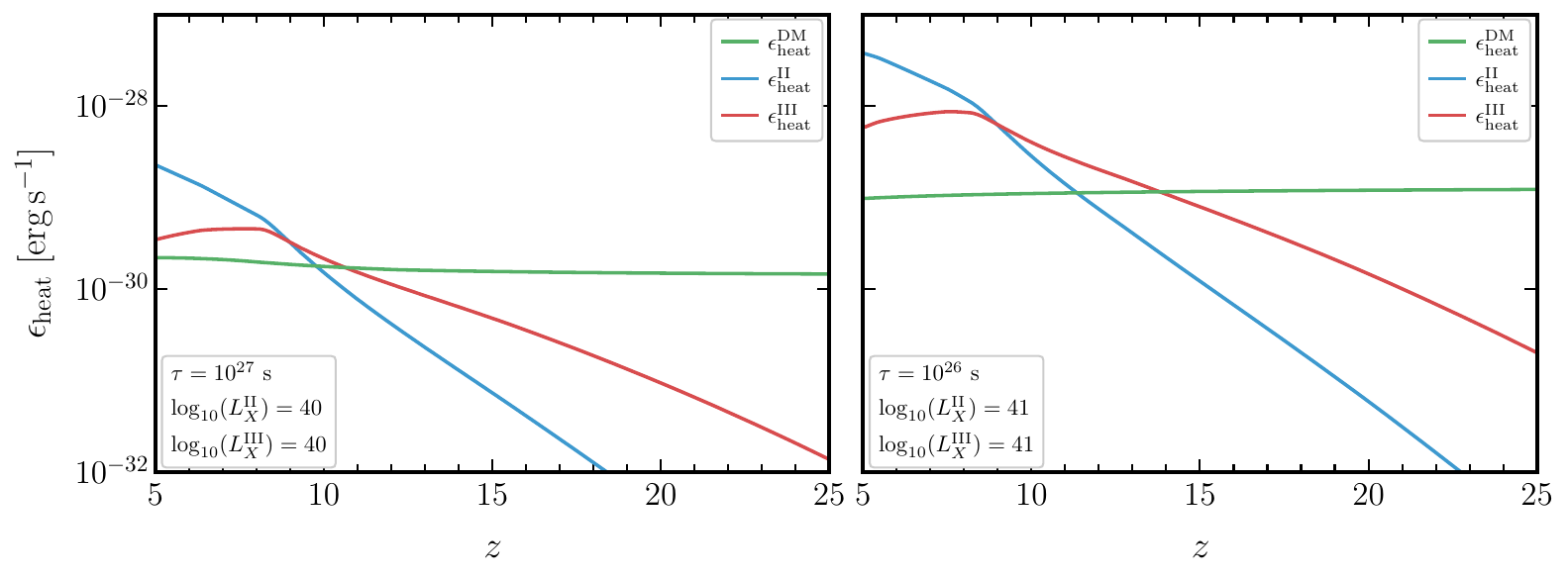}
\caption{Heating rates per baryon from \popII{}-dominated ACGs (blue), \popIII{}-dominated MCGs (red) and DM (green). For this figure, we have considered either $\log_{10}(L_X^{\rm II}) =\log_{10}(L_X^{\rm III}) = 40$ (left panel) or $\log_{10}(L_X^{\rm II}) =\log_{10}(L_X^{\rm III}) = 41$ (right panel), with the remaining parameters fixed to their fiducial values given in \reftab{tab:params}. The DM contribution is computed assuming a 100 MeV$/c^2$ DM particle species decaying into electron positron pairs with a lifetime $\tau = 10^{27}~{\rm s}$ (left panel) and $\tau = 10^{26}~{\rm s}$ (right panel).}
\label{fig:heating_rates}
\end{figure}

As heating will play an important role in this analysis, we illustrate different contributions to    the  heating rate per baryon, $\epsilon^\beta_{\rm heat}$ in \refequ{eq:TK} in \reffig{fig:heating_rates}. The blue and red curves show the X-ray heating rate from  ACGs and MCGs while the green curve is due to heating from DM decays discussed in \refsec{sec:DM_decay}. We assume that DM particles have mass $m_\chi = 100~{\rm MeV}/c^2$ and decay into electrons and positrons with a lifetime $\tau = 10^{27}~{\rm s}$  in the left panel and $\tau=10^{26}~{\rm s}$ in the right panel. We further show the impact of the value of $\log_{10}(L_X^{\rm II/III})$ fixing these parameters to 40 in the left panel or 41 in the right panel. The rest of the astrophysical parameters are set to the fiducial values listed in \reftab{tab:params}. By definition, MCGs  start to efficiently heat the medium before ACGs. However, due to rapid growth of the halo mass function, ACGs soon become dominant (see also, for example,  \cite{Qin2020MNRAS.495..123Q,Munoz2022MNRAS.511.3657M}). The DM contribution shows a flat dependence in redshift. Indeed, in the case of DM decays, the heating rate per baryons essentially scales as  $(1+z)^3/n_b(z)$, which is redshift independent, see eqs.~(\ref{eq:dep}) and ~(\ref{eq:dEdVdtinj}).  The prefactor $f_{\rm heat}(z,x_e)$ induces an extra  redshift dependence. The latter is rather mild for a  DM candidate decaying into electron-positron pairs between $z=5$ and  $25$. Similar redshift dependence is expected for low mass DM decaying into a pair of photons.
DM heating appears thus to be dominating at the earliest times.

The imprint of DM decay on the 21cm power spectrum is strongest at early times, $z\gtrsim10-15$ for these examples.

\subsection{Imprint in the 21cm signal}
\label{sec:DMimprint}

\begin{figure}
\centering
\includegraphics[width=0.49\linewidth]{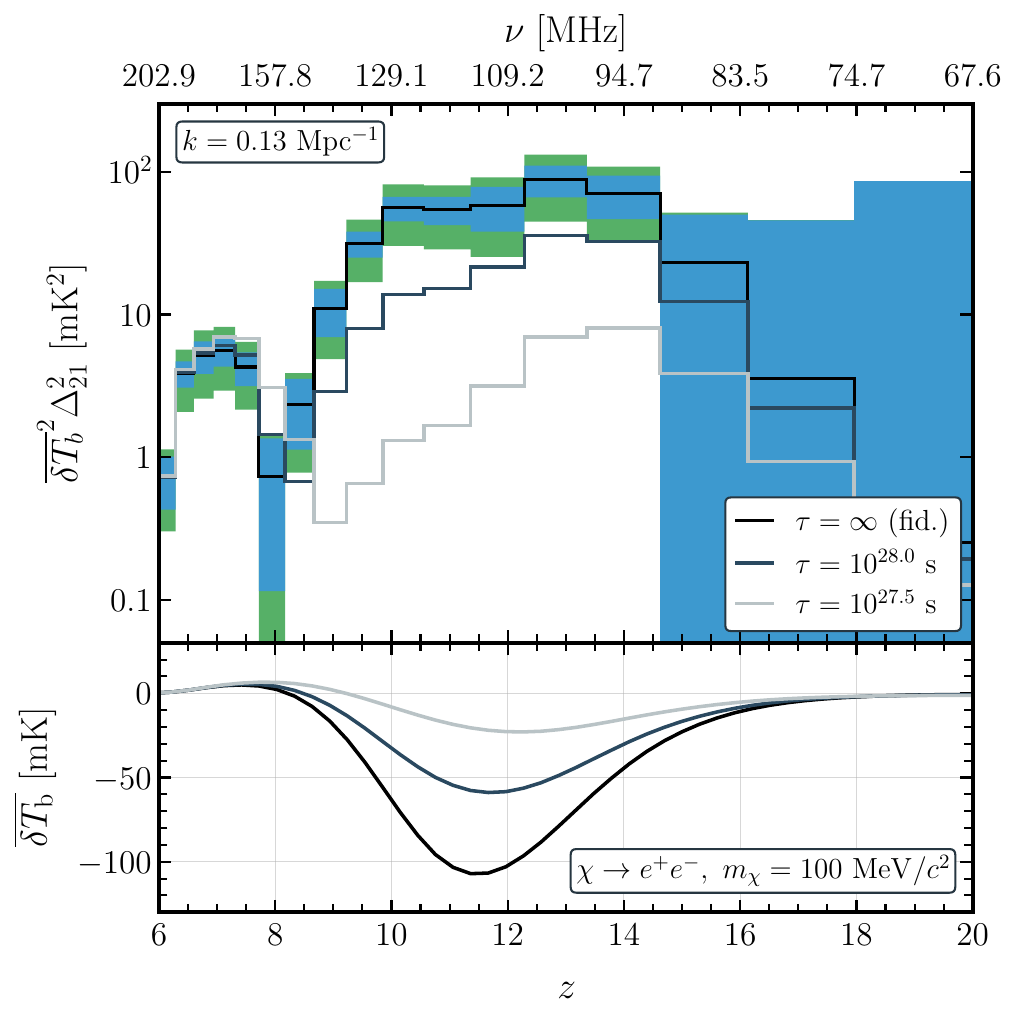}
\includegraphics[width=0.49\linewidth]{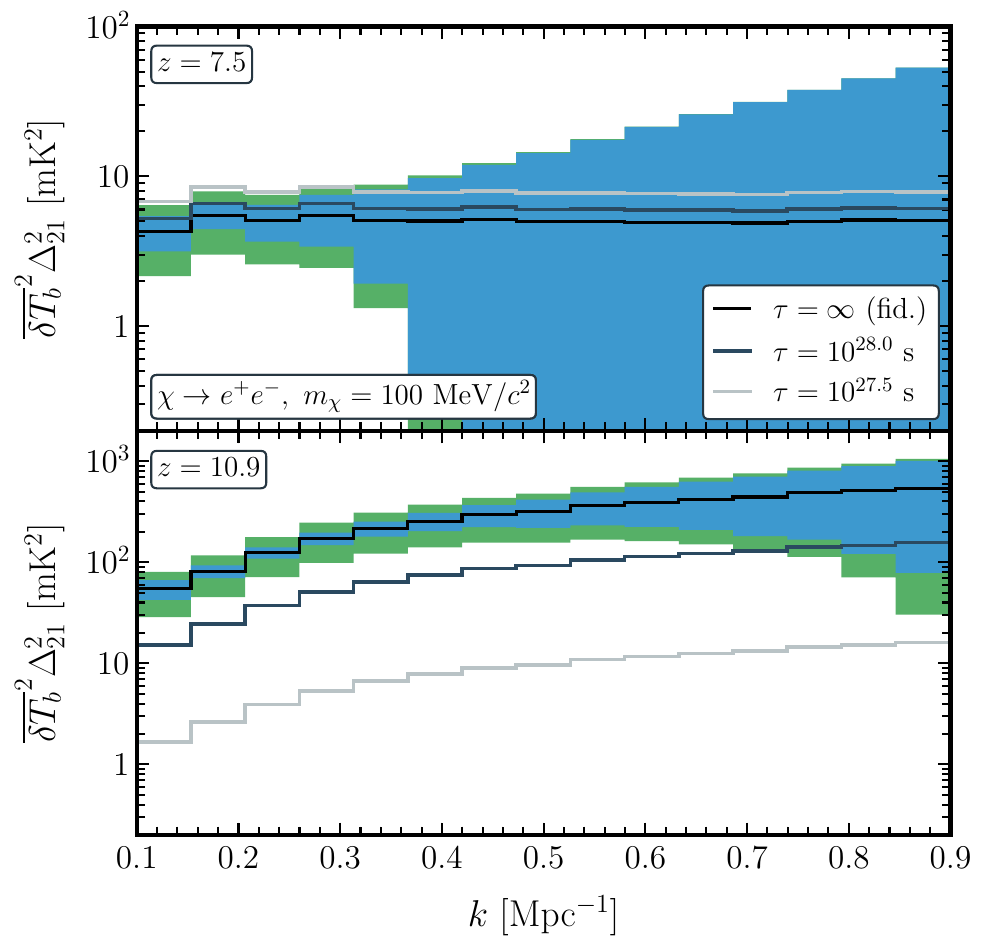}
\caption{Same as \reffig{fig:global-PS-LX} with the two gray lines illustrating the case of a 100 MeV$/c^2$ dark matter candidate decaying to electron positron pair with a lifetime of $10^{28}$ s (dark gray) and $10^{27.5}$ s (light gray). }
\label{fig:global-PS-LX-DM}
\end{figure}

As previously pointed out, dark matter energy injection mainly affects the 21cm signal as a new global heating source (e.g., refs.~\cite{Valdes:2009cq,Evoli:2014pva}). For the fiducial model considered here, the ionizing photons can not compete with the ones from astrophysical source.\footnote{As discussed in~\cite{Lopez-Honorez:2013cua, Poulin:2015pna,Liu:2016cnk}, CMB data prevent annihilating DM to be the dominant source of reionization while for low mass ($<100$ MeV/$c^2$) decaying DM contributions up to 10\% might be allowed. We further confirm that even assuming the shortest allowed DM lifetime from \reffig{fig:decaybound}, the DM contribution to ionization result in $x_e \lesssim 10^{-2}$ at all redshifts.} Figure \ref{fig:global-PS-LX-DM} illustrates the effect of this new source of heating on the global signal and the power spectrum when considering a 100 MeV$/c^2$ dark matter candidate decaying at 100\% into a electron positron pair with a lifetime of $10^{27.5}$ (light gray) and $10^{28}$ s (dark gray).  As visible in \refequ{eq:dEdVdtinj} a shorter lifetime implies a stronger heating of the IGM and hence induces a shallower absorption trough in the global signal.

Notice that \reffig{fig:global-PS-LX-DM} illustrates the case of a single population of galaxies (AGCs only). This is the scenario in which the DM heating can be more easily disentangled from the astrophysics sources as AGCs heating rate becomes comparable to the DM one at rather late times ($z\lesssim 10-15$), see \reffig{fig:heating_rates}.  In addition this  scenario is better constrained by  data with a minimal number of sources of heating and the least number of astrophysics  parameters that could be degenerate with DM heating, see~\refsec{sec:results}. The case of a  more complete, yet more complex,  astrophysics model is illustrated in~\refapp{app:MCGs} where we consider both ACGs and MCGs sources of heating. The latter scenario involves more sources of uncertainties  as MGCs properties are yet to be determined. We will also consider this scenario in our analysis.  MGCs can heat the medium earlier than AGCs and the DM imprint becomes less easy to untangle from astrophysics. This is expected to mitigate the constraints on exotic sources of heating, see~\refsec{sec:results} for a quantitative result. 

In the model illustrated in \reffig{fig:global-PS-LX-DM}, DM energy injection dominates IGM heating at early times when $z>10$--15. As a result, DM decays give rise to a  more uniformly heated IGM at early times, which can decrease the large-scale 21cm power during the EoH compared with galaxy-only heating (see also \cite{Evoli:2014pva,Lopez-Honorez:2016sur}.  On the other hand, the impact of DM decay on the late-time 21cm power ($z<9$) is far more modest.  As mentioned earlier, the contribution to reionisation from DM decay is sub-dominant to that from stars, even for even the shortest lifetimes considered here.  Our shortest lifetime model (gray curve) does show a slight enhancement of 21-cm power in the early stages of the EoR ($z\sim$ 7--8), when the signal is in emission. The additional heating from DM decay increases the $(1 - T_{\rm CMB}/ T_S)$ temperature term from \refequ{eq:deltaTb}.  However, this term quickly saturates to unity (due to the X-rays from stars) and so the different DM lifetimes considered here do not impact the 21-cm power below $z \lesssim 7$.

\begin{figure}
\centering
\includegraphics[width=0.95\linewidth]{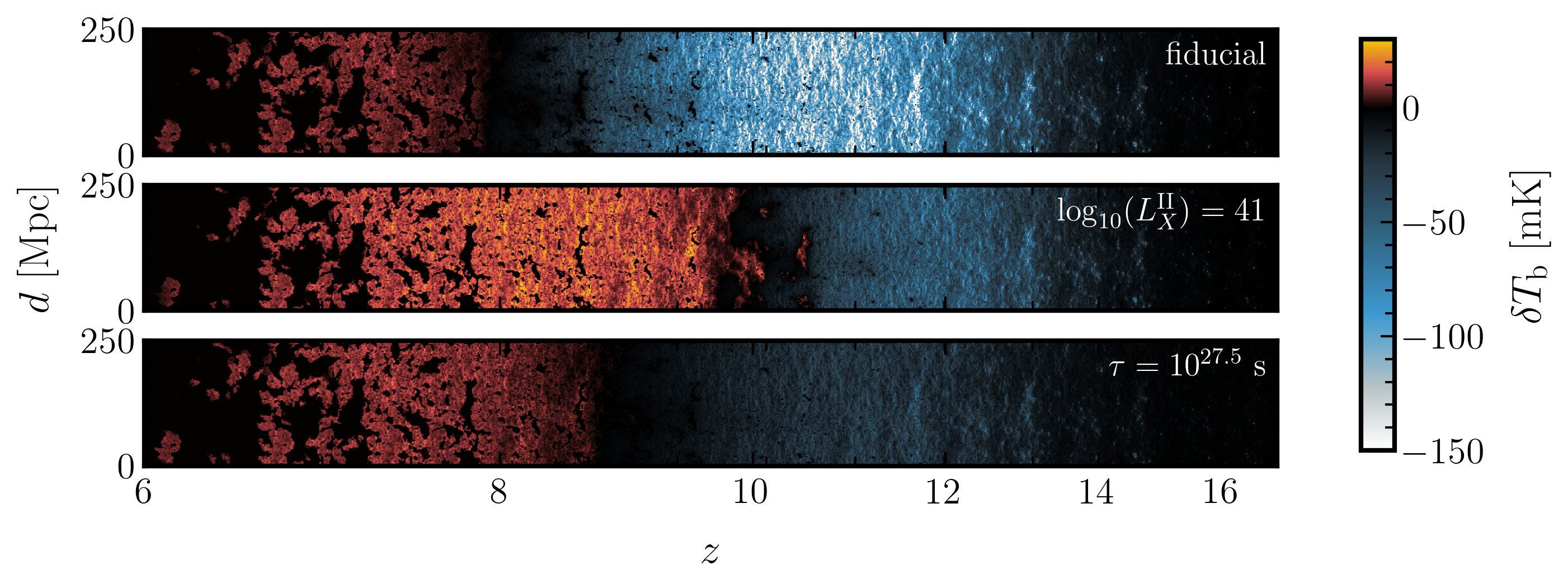}
\caption{Lightcone slices of the differential brightness temperature in our $(250~{\rm Mpc})^3$ large simulation box. We show results for the fiducial model with $\log_{10}(L_X^{\rm II}) = 40$ and $\tau = \infty$ (upper panel), $\log_{10}(L_X^{\rm II}) = 41$ (middle panel), and for a 100 MeV$/c^2$ decaying DM with $\tau = 10^{27.5}~{\rm s}$ (bottom panel).}
\label{fig:lightcones}
\end{figure}

The discussion above is further illustrated with \reffig{fig:lightcones} that shows  a 2D slice through the brightness temperature lightcone. We show the fiducial model with $L_X^{\rm II}=10^{40}$ and no decaying DM (i.e., $\tau=\infty$) (top panel), a model with larger $L_X^{\rm II}$ (central panel) and a model with a shorter lifetime $\tau$ (bottom panel). The figure illustrates how the different features of the power spectrum (in Fourier space) relate to the real space signal -- e.g., a lower power spectrum is related to a weaker signal and/or low contrast. 

We conclude that heating from DM decay has a qualitatively similar imprint on the 21cm signal as heating from galaxies, but their redshift and scale dependencies can be very different. Given that experiments such as HERA will be able to probe a large range of redshifts and scales (see \refsec{sec:21cm_physics}), we expect them to be able to disentangle these two sources of heating. From the sensitivity estimates shown in \reffig{fig:global-PS-LX-DM}, we expect HERA could probe lifetimes up to $~10^{27}-10^{28}$ s, surpassing the sensitivity from CMB and Lyman-$\alpha$ probes.  We quantify this further below.

\section{Method and results}
\label{sec:method}

Our goal is to forecast the sensitivity of the full HERA array with 331 antennas to DM decay. For that purpose multiple approaches are possible. A cost-efficient solution in terms of computational resources is the Fisher Matrix. For a given  set of cosmological parameters $\boldsymbol{\theta}$ the Fisher information matrix is defined as
\begin{equation}
    F_{ij} \equiv -\left\langle\frac{\partial \ln {\cal L}}{\partial\theta_i\partial\theta_j}\right\rangle
\end{equation}
where $\cal L$ denotes  the likelihood, the expected distribution of the data given a certain model. 
The Cr\'amer-Rao theorem \cite{frechet1943extension, darmois1945limites, aitken1942xv} states that the marginalized error, $\sigma_{\theta_i}$, on a given parameter $\theta_i$ follows $\sigma_{\theta_i} \geq \sqrt{(F^{-1} )_{ii}} $, implying the Fisher matrix approach always gives a local optimistic estimate of the errors. In the following, we estimate the expected variance from the diagonal elements of the inverse Fisher Matrix components,
\begin{equation}
    \sigma_{\theta_i}^2 = (F^{-1} )_{ii} \, .
\label{eq:CReq}
\end{equation} 
More details on the Fisher matrix method are given in \refapp{app:Fisher}.

In the following, we describe our treatment of the Fisher Matrix forecasts within a new automatized python package called \cmCAST{}\footnote{\url{https://github.com/gaetanfacchinetti/21cmCAST}} -- based on  \cmfish{}~\cite{Mason:2022obt}. We then discuss the results of our analysis, which demonstrates that 21cm cosmology could provide key constraints on dark matter energy injection.

\subsection{Fisher matrix analysis}
\label{sec:method:Fisher_matrices}

\begin{table}[t!]
    \begin{tabular}{c c | c c c c c}
         \multirow{2}{*}{ACGs} & param. &  $\log_{10}(f_{\star, 10}^{\rm II})$ &  $\alpha_\star^{\rm II}$ & $\log_{10}(f_{\rm esc, 10}^{\rm II})$ & $\log_{10}(L_{X}^{\rm II})$ &   \\
         \hhline{~------}
         & fiducial & -1.3 & 0.5 & -1.0 & 40.0 & \\[5pt]
         \multirow{2}{*}{MCGs} & param. &  $\log_{10}(f_{\star, 7}^{\rm III})$ &  $\alpha_\star^{\rm III}$ & $\log_{10}(f_{\rm esc, 7}^{\rm III})$ & $\log_{10}(L_{X}^{\rm III})$ \\
         \hhline{~-----}
         & fiducial & -2.0 & 0.5 & -2.0 &  40.0 &  \\[5pt]
        & param & $t_\star$ & $\alpha_{\rm esc}$  & $\frac{E_0}{\rm keV}$ &$\log_{10}({M_{\rm turn}}/{ M_\odot})$  \\
        \hhline{~-----}
        & fiducial & 0.5 & -0.5 & 0.5 & 8.7& 
    \end{tabular}
    \caption{Values of the astrophysical parameters in the fiducial models. Parameters specific to \popII{}-dominated ACGs and \popIII{}-dominated MCGs are listed in the first and second row, respectively. The other parameters are in the last row. Note that ${M_{\rm turn}}$ is only used in the simplified one-population model (see more details in \refsec{sec:starform}).}
    \label{tab:params}
\end{table}

We evaluate the 21cm power spectrum $\overline{\delta T_{\rm b}}^2\Delta_{21}^2$ on a  fixed grid of modes $k$ and redshifts $z$. We assume a total bin number of $N_{k} \times N_z$ and consider uncorrelated $z$ and $k$ bins as in~\cite{Mason:2022obt}. Assuming that the likelihood $\cal L$ can be described by a Gaussian distribution for each bin, the Fisher matrix elements reduce to the sum
\begin{equation}
    F_{ij} = \sum_{i_k}^{N_k}\sum_{i_z}^{N_z} \frac{1}{\sigma_\Delta^2(k_{i_k}, z_{i_z})}\frac{\partial \overline{\delta T_{\rm b}}^2(z_{i_z} \, | \, \boldsymbol \theta)\Delta_{21}^2(k_{i_k}, z_{i_z} \, | \, \boldsymbol \theta)} {\partial \theta_i} \frac{\partial \overline{\delta T_{\rm b}}^2(z_{i_z} \, | \, \boldsymbol \theta)\Delta_{21}^2(k_{i_k}, z_{i_z} \, | \, \boldsymbol \theta)} {\partial \theta_j} \, .
    \label{eq:FisherMatrix}
\end{equation}
In accordance with the common practice in the field \cite{Park:2018ljd, Mason:2022obt, Zahn:2010yw}, we account for three contributions to the measurement error in the power spectrum:
\begin{equation}
    \sigma_\Delta^2 \equiv \sigma_{\rm exp}^{2} + \sigma_{\rm Poisson}^2 + \left[0.2 \overline{\delta T_{\rm b}}^2\Delta_{21}^2\right]^2\, .
    \label{eq:sigmaDelta}
\end{equation}
The first contribution, $\sigma_{\rm exp}$ denotes the experimental error (i.e., thermal noise), which is evaluated using the public code {\tt 21cmSense} \cite{2013AJ....145...65P,2014ApJ...782...66P}\footnote{\url{https://github.com/jpober/21cmSense}}. In that code we fix the HERA experiment as an hexagonal array of 331 antennas (11 on each side) with separation and dish size of 14~m, located at a latitude of $\sim 30.8^\circ$. Moreover, we fix the bandwidth to $B=8~{\rm MHz}$ and the spectral resolution to $\delta \nu \sim  100~{\rm kHz}$. We assume a total operating time of 1000 hrs (6 hrs per day during 166.7 days). We adopt the \emph{moderate} foregrounds setting, with the default system temperature for the 21cm line, implying a system temperature with the following redshift dependence:\footnote{Notice that  \refequ{eq:Tsys}, corresponds to the \emph{pessimistic} scenario of \cite{Mason:2022obt}. }
\begin{equation}
    T_{\rm sys}(z) = 100~{\rm K} + 260~{\rm K}\left(\frac{\nu(z)}{150 ~{\rm MHz}}\right)^{-2.6} \sim   100~{\rm K} + 300~{\rm K}\left(\frac{1+z}{10}\right)^{2.6} \, ,
    \label{eq:Tsys}
\end{equation}
with $\sigma_{\rm exp}(k_{i_k}, z_{i_z})\propto T_{\rm sys}^2(z_{iz})$\cite{2013AJ....145...65P}.  For illustration, this thermal noise is shown in blue in Figs.~\ref{fig:global-PS-LX} and~\ref{fig:global-PS-LX-DM}. The second contribution to the measurement error in \refequ{eq:sigmaDelta} is the cosmic variance, $\sigma_{\rm Poisson}$, arising from the evaluation of the power spectrum in a finite size simulation box. Finally, as a third contribution to $\sigma_{\Delta}$, we consider a modeling uncertainty of 20\% \cite{Zahn:2010yw}. For illustration, the total measurement error is shown  in green in  figures~\ref{fig:global-PS-LX} and~\ref{fig:global-PS-LX-DM}. 

As shown in \refequ{eq:Tsys}, the thermal noise increases with the redshift.  We have set the maximum redshift considered in this analysis to $z_{\rm max} \sim 20$, after having checked that higher redshifts have a negligible signal-to-noise ratio, and do not impact our results.  In addition, because at low redshifts reionization sharply suppresses the signal, we set the minimum redshift to $z_{\rm min}=6$. The bin size in redshift is set by the bandwidth $B$. Because foregrounds increasingly contaminate large scale modes (e.g., \cite{2013AJ....145...65P}), we set the minimum $k$ value to  $k_{\rm min} = 0.1~{\rm Mpc^{-1}}$ and the bin size to $\Delta k_{\parallel}(z_{\rm min}, B) = 0.053~{\rm Mpc}^{-1}$. In addition, since the thermal noise drastically increases at large $k$, we further restrict the bin range to $k_{\rm max} = 1~{\rm Mpc^{-1}}$.  We check that our results are not sensitive to this choice.
For the exact values used in our analysis, see tables~\ref{tab:zkbins}. \\

Adopting a similar approach as \cmfish{}~\cite{Mason:2022obt}, {\tt 21cmCAST} generates a series of configuration files in which all parameters of the model are varied around their fiducial values, $\boldsymbol \theta_{\rm fid}$,  by a few percent so as to numerically calculate the derivatives of the power spectrum.  The configuration files are then given as input for \cmFAST{} which produces \emph{lightcones} of the corresponding simulated universe. Here we work with simulation boxes of size 250~Mpc and divided into $128^3$ cells. Eventually, \cmCAST{} gathers all the \emph{lightcones}, bins  the power spectra, evaluates the experimental noise using {\tt 21cmSense}, computes  the derivatives numerically  and outputs the Fisher matrix. The main differences between \cmCAST{} used here and \cmfish{} are: \emph{(i)} more flexibility with the binning choice and \emph{(ii)} the complete integration of {\tt 21cmSense} to evaluate the noise directly from the chosen fiducial model. \\

Through this Fisher matrix analysis, we evaluate the best lower limits that are expected to be set on the dark matter lifetime $\tau$ by the HERA experiment within two possible astrophysics scenarios and fixed dark matter mass $m_\chi$.
The first scenario considers only radiation from \popII{}-dominated ACGs. Unlike MCGs, ACGs have been observed and their stellar to halo mass relation is well constrained by UV luminosity functions.  In this case, the minimal set of astrophysical (with one single population of galaxies) and dark matter parameters considered in our Fisher Matrix analysis is:
\begin{equation}
   \boldsymbol{\theta}_{\rm II} =\{    \log_{10}(f_{\star, 10}^{\rm II}), \alpha_\star^{\rm II}, t_\star,\log_{10}(f_{\rm esc, 10}^{\rm II}), \alpha_{\rm esc}, \log_{10}(M_{\rm turn} / M_\odot), \log_{10}(L_{X }^{\rm II}), E_0, \Gamma\}  \,, 
   \label{eq:theta}
\end{equation}
i.e., nine parameters with only the last one, the DM decay rate $\Gamma$, that sets the DM decay contribution with lifetime $\tau=1/\Gamma$.
 
We also consider an extended model that includes an additional putative contribution from unseen MCGs.  This model is parametrized by:
    \begin{eqnarray}
         \boldsymbol{\theta}_{\rm II + III}&=&\{ \log_{10}(f_{\star, 10}^{\rm II}), \alpha_\star^{\rm II}, t_\star,\log_{10}(f_{\rm esc, 10}^{\rm II}), \alpha_{\rm esc}, \log_{10}(L_{X }^{\rm II}), E_0, \Gamma,\,\nonumber \\
         & &\hspace{.3cm} \log_{10}(f_{\star, 7}^{\rm III}), \alpha_\star^{\rm III}, \log_{10}(f_{\rm esc, 7}^{\rm III}), \log_{10}(L_{X }^{\rm III}  )    \} \,. 
   \label{eq:thetamini}
    \end{eqnarray}
i.e., 12 parameters characterizing PopII-dominated ACGs and PopIII-dominated MCGs\footnote{Notice that $\log_{10}(M_{\rm turn} / M_\odot)$ is not considered as a parameter in $\boldsymbol{\theta}_{\rm II+III}$ as, within the ACG \& MCG two population approach, the threshold masses are computed according to photoheating and LW background assumed in the analysis, see more details in \refsec{sec:starform}.}. Because MCGs appear before ACGs, their contribution to the 21-cm PS would be more degenerate with that of DM decay (see \reffig{fig:heating_rates}). The bounds on the DM decay properties obtained from $\boldsymbol{\theta}_{\rm II+III}$ should then be more conservative than from $\boldsymbol{\theta}_{\rm II}$. The fiducial values considered in this work are tabulated in \reftab{tab:params}. 

Notice that the constraint on the DM lifetime $\tau$, or equivalently its decay rate $\Gamma$, is derived assuming a fiducial model with no exotic energy injection. We thus have a fiducial value of the decay rate set to $\Gamma_{\rm fid} = 0$, i.e., $\tau=\infty$. More details on the treatment of this parameter in our Fisher analysis, which comes with some technical difficulties,  is provided in  \refapp{app:Fisher}.

\subsection{Results}
\label{sec:results}

\begin{figure}[t!]
\centering
\includegraphics[align=c, width=0.45\linewidth]
{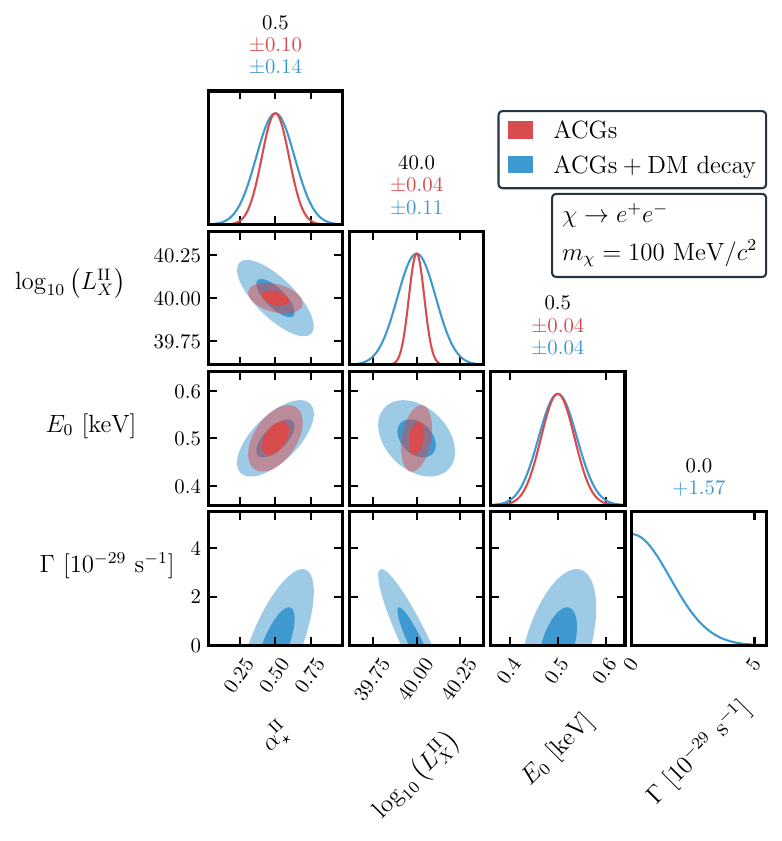}
\includegraphics[align=c, width=0.53\linewidth]
{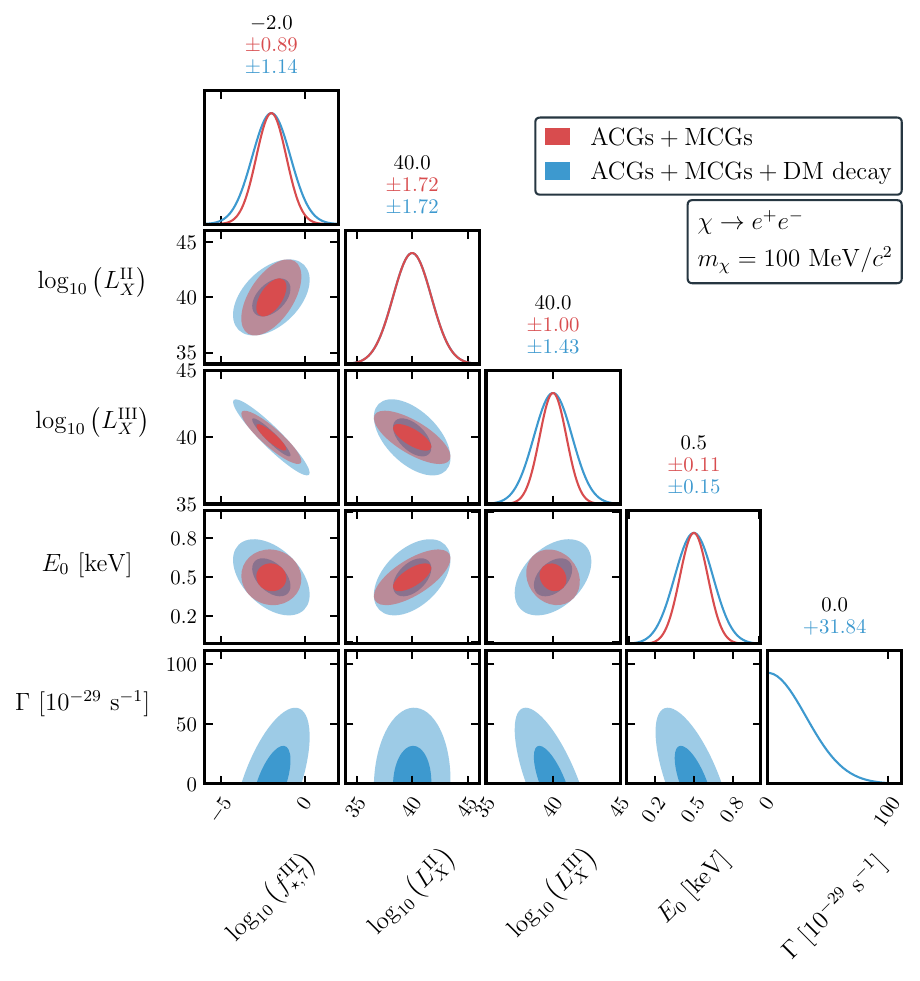}
\caption{{\small Marginalized  posterior from our Fisher matrix analysis for  reduced sets of parameters.  The blue contours corresponds to forecasts for a model with   100 MeV$/c^2$ DM decaying into electron positron pairs. For comparison, we show in red the results of our Fisher forecasts without DM decays. The top right plots show 1D marginalized posteriors and the lower triangle plots show the 2D marginalized posteriors. In the 2D plots, we show with dark (light) colors the    1 (2) $\sigma$ confidence intervals for the posteriors. Above the top right plots, fiducial values of the parameters are given in black while the 1 sigma errors with (without) decaying dark matter quoted in blue (red). The left (right) panel shows our results for $\boldsymbol{\theta}_{\rm II}$ (for $\boldsymbol{\theta}_{\rm II +III}$). For completeness, the full triangle plots are provided in~\refapp{app:Fisher_triangle_plots}.
}}
\label{fig:Triangle_Zoom}
\end{figure}

Triangle plots resulting from our Fisher matrix analysis for the set of parameters $\boldsymbol{\theta}_{\rm II}$ and $\boldsymbol{\theta}_{\rm II + III}$  are shown in the left and right panel of \reffig{fig:Triangle_Zoom} respectively. More precisely we show in both cases the marginalised posterior distributions of the parameters that are most degenerate with the DM decay rate $\Gamma$ (assuming that they are Gaussian). See \refapp{app:Fisher_triangle_plots} for the full triangle plots with all parameters in $\boldsymbol{\theta}_{\rm II}$ and $\boldsymbol{\theta}_{\rm II + III}$. Here we consider decay into electron positron pairs and a DM mass fixed to $m_\chi=100$ MeV$/c^2$. Similar plots can be obtained for different DM masses or decays into photons. The dark (light) blue contours  represent the 1- (2-)sigma confidence intervals of the two-dimensional marginalized posterior probability distributions while, in the right-most plots, the blue lines are the one dimensional marginalized posteriors for the full set of cosmological parameters $\boldsymbol{\theta}_{\rm II}$ and $\boldsymbol{\theta}_{\rm II+III}$. Above the latter plots we provide the fiducial values of the parameters (in black), $\theta_{i,\, \rm fid}$, and the corresponding one-sigma error (in blue), $\sigma_{\theta_i}$.  

To assess the validity our Fisher-matrix analysis, we also provide the posteriors and the estimated 1-sigma errors (in red) 
that we obtain when the decay parameter, $\Gamma$, is not considered in the analysis. The corresponding curves can easily be compared to the results of the Fisher Matrix analysis of~\cite{Mason:2022obt}. Our estimated errors for the common set of parameters are in very good agreement. 

\begin{figure}
    \centering
    \includegraphics[width=0.99\linewidth]{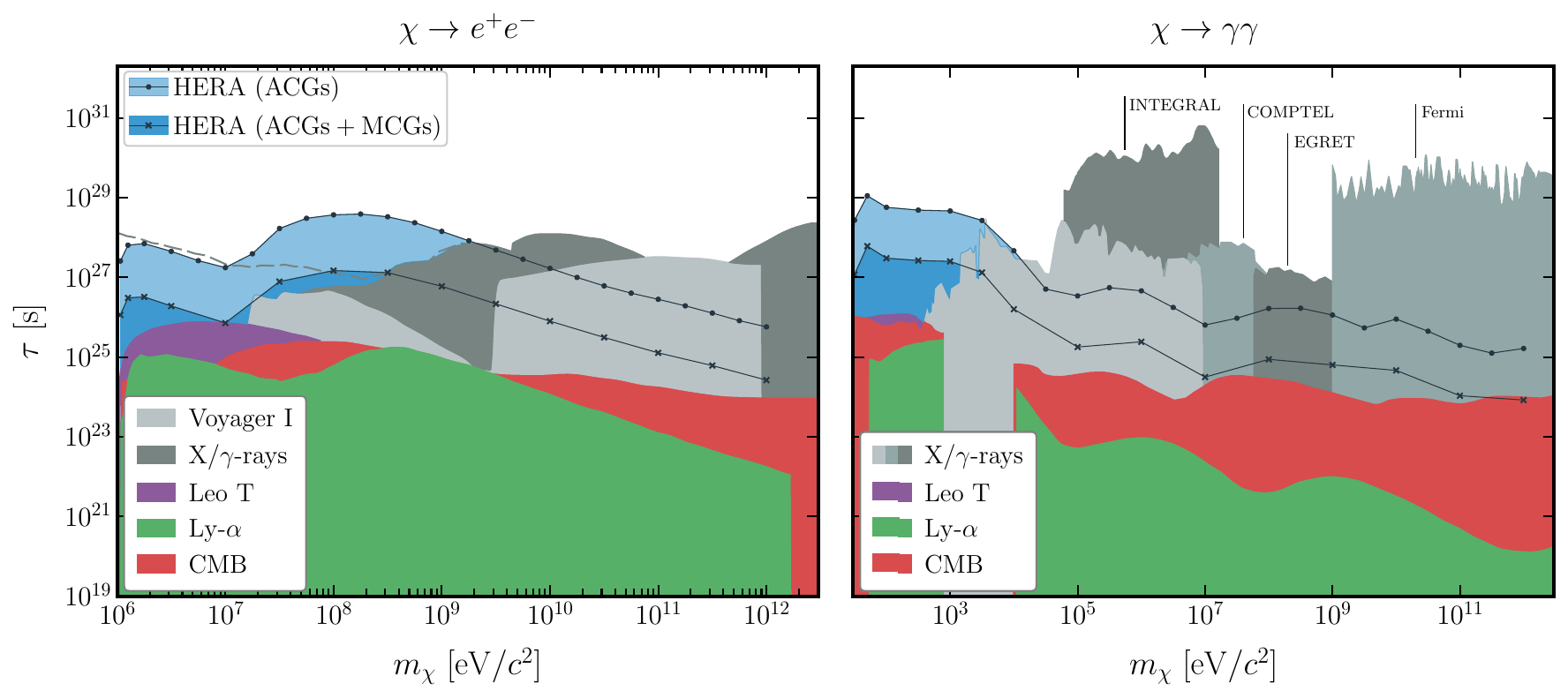} 
    \caption{ \small  Compilation of constraints on the dark matter lifetime (at 95\% level) for decay into an electron/positron pair (left panel) and photons (right panel). We superpose the forecasts for the HERA telescope assuming \popII{}-dominated ACGs only (light blue area, round markers) or \popII{}-dominated ACGs  + \popIII{}-dominated MCGs (dark blue, cross markers) with existing constraints. Green and red areas show the cosmological bounds set by Lyman-$\alpha$ forest~\cite{Liu:2020wqz, Capozzi:2023xie} and CMB~\cite{Slatyer:2016qyl, Capozzi:2023xie} data. Gray areas show astrophysical constraints. In the left panel we have reported the bound from the Voyager I observation of cosmic rays \cite{Boudaud:2016mos, Boudaud:2018oya} (light gray) and the constraint from X-/$\gamma$-ray experiments \cite{Essig:2013goa, Cohen:2016uyg, Massari:2015xea, Cirelli:2023tnx} (dark gray). In the right panel we show the X-ray limit from~\cite{Cadamuro:2011fd} in light gray colors, as well as bounds from  INTEGRAL/SPI \cite{Calore:2022pks}, COMPTEL, EGRET \cite{Essig:2013goa} and Fermi \cite{Foster:2022nva} with different shades of gray. The purple areas show the conservative constraints from the Leo T dwarf galaxy \cite{Wadekar:2021qae}. The dashed grey line on the left panel shows the latest XMM-Newton constraint with an improved treatment of cosmic ray propagation \cite{DelaTorreLuque:2023olp}.}
    \label{fig:decaybound}
\end{figure}

The marginalized error on $\log_{10}(L_{X }^{\rm II})$ increases by a factor of $\sim 3$ when introducing $\Gamma$, for our ACG only model. This is expected from the discussion in Secs.~\ref{sec:21numerics} and~\ref{sec:DMimprint} as both galaxies and DM decay play an important role in the evolution of the IGM temperature in the redshift range of interest. Understandably, we see that $\Gamma$ and $\log_{10}(L_{X}^{\rm II})$ are anti-correlated. $\Gamma$ is also degenerate, although less strongly, with $\alpha_\star^{\rm II}$ and $E_0$. Indeed, in those cases the marginalized 1-D posteriors are much more weakly affected and the 1-sigma error changes by at most $30\%$. As visible in \refequ{eq: LX}, increasing $E_0$ makes the X-ray photon spectrum heating the IGM harder.  Harder X-rays have longer mean free path (MFP). If the MFP is sufficiently long and heating happens in a highly homogeneous manner, then lowering $E_0$ leads to more efficient heating. Indeed more photons have MFP shorter than the Hubble length and can be absorbed in the IGM during the EoH. In the latter case, we need less efficient heating from DM, i.e., $E_0$ and $\Gamma$ are positively correlated as in  the left panel of~\reffig{fig:Triangle_Zoom}. On the other hand, from \refequ{eq:fstar}, we see that increasing $\alpha_*$ penalizes the star formation rate (SFR) of small mass halos that would contribute more to an early heating of the IGM than large mass halos. It is thus positively correlated with $\Gamma$ that induces an early heating. 

The parameters with the strongest degeneracies with $\Gamma$ in  the extended parameter space $\boldsymbol{\theta}_{\rm II+III}$ are shown in  the right panel of \reffig{fig:Triangle_Zoom}. This time it is the posterior of the X-ray normalisation factor associated to the \popIII{}-dominated MCGs ($\log_{10}(L_{X }^{\rm III})$), instead of that of the \popII{}-dominated ACGs ($\log_{10}(L_{X}^{\rm II})$), that is more strongly affected by the introduction of decaying DM.   This is to be expected as, by definition, \popIII{}-dominated MCGs form earlier than \popII{}-dominated ACGs. Thus,  MCGs contribute earlier to IGM heating, similar to decaying DM. Because our $\log_{10}(L_{X }^{\rm III})$ normalization is defined as the X-ray luminosity above $E_0$, lowering $E_0$ at a fixed value of $\log_{10}(L_{X }^{\rm III})$ decreases the luminosity at the highest X-ray energies that heat the IGM homogeneously.  This decrease in the galactic homogeneous IGM heating can be compensated by increasing the (homogeneous) DM decay heating, i.e., increasing $\Gamma$. $E_0$ and $\Gamma$ are thus anticorrelated as seen in the right panel of \reffig{fig:Triangle_Zoom}. On the other hand, higher $f_{*,7}^{\rm III}$ implies stronger feedback which could eventually suppress the low-mass ACGs at lower redshifts. This again suppresses homogeneous heating from stars and can be compensated by larger DM heating.\footnote{No such correlation appear in the AGCs-only case, see \reffig{fig:Triangle}, as the feedback mechanisms were neglected in that case.}

When considering the extended parameter space, the 1-sigma upper bound on $\Gamma$ is weaker by approximately one order of magnitude when fixing the decaying DM mass to 100 MeV$/c^2$. Indeed,  when considering the minimum set $\boldsymbol{\theta}_{\rm II}$, the heating effect of ACGs like galaxies only, which appear relatively late compared to the heating of DM, and the wide redshift range probed by HERA allows to disentangle  the DM effect from the astrophysics and sets strong constraints on the DM decay rate. In the case of MCGs+ACGs, with MCGs heating the IGM at earlier times than ACGs, the effect of DM is more easily drawn by the effect of the astrophysics parameters and the limits on the decay rate are much weaker. 

Our main results, presented in \reffig{fig:decaybound}, showcase the lower bound at a 95\% confidence level (CL) on the lifetime of dark matter (DM) derived from our Fisher matrix forecasts based on HERA specifications.  Black lines with bullets are obtained for one single population of galaxies while  black lines with crosses assume ACGs+MCGs. The blue area below these curves are excluded at 95\% CL. As expected, when considering the ACGs+MCGs  scenario,  heating from galaxies competes with DM heating earlier and the DM heating parameters become more difficult to constrain.  The lower bound on the DM life time  becomes thus less stringent in ACGs+MCGs scenario (crosses) than in the AGCs only case (bullets).

More precisely, our bounds are actually obtained conducting a Fisher matrix analysis at each value of the DM mass marked with a cross or a bullet. We show the results both for decays into electron positron pairs (left panel) and photons (right panel). The 95\% CL bound just simply corresponds to $\tau (m_\chi)< (2\times\sigma_\Gamma(m_\chi))^{-1}$, where $\sigma_\Gamma$ is obtained saturating the Cr\'amer-Rao inequality as in \eqref{eq:CReq}. It is thus an optimistic limit.\footnote{In \cite{Mason:2022obt}, it was observed that the 1-$\sigma$ credible intervals obtained from a Fisher matrix  analysis  on a given  set of parameters are typically within 40\%  of the ones resulting from a MCMC analysis.}  For 
decays into $e^+e^-$, a few 100 MeV$/c^2$ DM gets the most stringent 21cm  bounds on  $\Gamma$ while, for decays into photons, it is the case for the lowest DM masses (with $m_\chi<$MeV$/c^2$). The mass dependency  of our 21cm forecasts actually traces back to the dependency of energy deposition efficiencies, $f_a(x_e,z)$, on  the injected energy (see e.g.,~\cite{Slatyer:2012yq,Slatyer:2015jla,Slatyer:2015kla}).\footnote{Notice that we do not see any difference in our bounds including or not backreaction in the computation of $f_a(x_e,z)$. That is because the Fisher Matrix analysis probes only small variations around the fiducial where $\Gamma=0$, see the details in \ref{app:DMdecay}. For such small shifts in $\Gamma$, DM decays do not impact sufficiently  $x_e$ to change significantly  the deposition efficiencies. We expect though that employing a MCMC we shall be able to test the dark matter (DM) backreaction.}

In \reffig{fig:decaybound}, we  also show the lower bounds arising from cosmology probes such as Lyman-$\alpha$ forest~\cite{Liu:2020wqz} (green) or CMB~\cite{Slatyer:2016qyl,Capozzi:2023xie} (red).  21-cm measurements with HERA could improve by up to 3 orders of magnitude the current limits on the DM lifetime, when considering \popII{}-dominated ACGs only (black line with bullets). Note that mass dependency seen in the Lyman-$\alpha$ constraints (green area) is similar to the one of our 21cm bound forecasts (bullet or crosses).  This is because Lyman-$\alpha$  data also probe the DM heating and is thus sensitive to $f_{\rm heat}$ energy injection dependency.  Yet, these data probe the IGM temperature  at lower $z$, when $T_k$ is orders of magnitude larger, resulting into weaker constraints compared to our 21cm  forecast (for $f_{a}$ as a function of redshift and  energy injection see also e.g.~\cite{Slatyer:2015kla}). Even when \popIII{}-dominated MCGs (dark blue area, line with crosses) are included in our analysis, the 21cm  constraints still improve by a factor of 10 to 100 compared to existing cosmology constraints in the lower DM mass range. 

In \reffig{fig:decaybound}, we additionally compare our forecast against constraints from the Voyager I observation of cosmic rays~\cite{Boudaud:2016mos, Boudaud:2018oya} and the result of X- or $\gamma$-ray experiments~\cite{Cadamuro:2011fd,Essig:2013goa, Cohen:2016uyg, Massari:2015xea} including those coming from INTEGRAL/SPI \cite{Calore:2022pks}, COMPTEL, EGRET \cite{Essig:2013goa} and Fermi \cite{Foster:2022nva}. We find that existing constraints for dark matter heavier than 1 GeV$/c^2$ (or 100keV$/c^2$) for decays into $e^+e^-$  (or photons) remain competitive and our 21cm forecast for 1000 hours of HERA observation is unlikely to improve these limits in the higher DM mass range.

\section{Conclusion and future perspectives}
\label{sec:conclusion}

Determining the nature of DM is one of the major goals of particle physics and cosmology today. It has brought us to explore all possible ways to shed new light on its properties. Cosmology probes are well known to set stringent bounds on DM energy injection into the IGM. In this work, we have carefully studied the prospects for 21cm cosmology to probe dark matter decays into electron-positron pairs and photons. For that purpose, we have developed a new version of the public code \cmFAST{}, that interfaced with the public package \DarkHistory{}, accounts for DM energy deposition into heat, ionization and excitation of the medium and its effect on the 21cm signal. In particular, DM energy injection is expected to mainly affect the 21cm global signal and its power spectrum  as an exotic source of heating, already efficient before \popII{}-dominated ACGs light on.

As is well known, DM decays give rise to a relatively late time energy deposition into the medium. This makes late time probes, such as Lyman-$\alpha$ forest or 21cm cosmology, very interesting targets to detect the DM  imprint.  In this work, we focus on the effect of DM on the 21cm signal power spectrum and prospects for constraints on the DM lifetime by the HERA interferometer. This telescope will enable us to explore a vast range of redshifts, stretching from the Epoch of Reionization to Cosmic Dawn, with exceptional precision. This capability is of paramount importance because DM is not the sole contributor to the heating process, and it is crucial to distinguish its distinct signature from that generated by X-rays emitted from the first galaxies. In our work, we argue that their different imprint in the probed redshift range is the key to obtain competitive constraints with respect to existing probes, both from cosmology and from astro-particle physics experiments.

In order to provide quantitative forecasts, we have performed for the first time a dedicated Fisher Matrix analysis considering the HERA telescope and two different astrophysics scenarios including DM decays. For the Fisher matrix forecast we have developed our own tool, \cmCAST, that interfaces with {\tt 21cmSense} to evaluate the anticipated experimental errors from 1000 hours of observation using all 331 antennas from HERA, extracting the expected marginalized error on a set of astrophysics parameter of our choice.

Our results, summarized in \reffig{fig:decaybound}, are very promising. When considering the minimal astrophysics scenario, HERA is expected to  improve on existing  cosmology constraints (from CMB and Lyman-$\alpha$ probes)  on the DM lifetime by up to 3 orders of magnitude. We also compare these prospects to the case where the astrophysics model includes both  \popII{}-dominated ACGs and \popIII{}-dominated MCGs. Similarly to DM, MCGs give rise to a new source of IGM heating before \popII{}-dominated ACGs light on, partially drowning the DM signal. Nevertheless, even in the latter case, HERA can improve on  existing  cosmology constraints  by a factor of 10 to 100. Finally, compared to existing $\gamma$-ray and cosmic-ray limits, HERA is expected to be a key player in constraining DM candidates decaying to $e^+e^-$ in the mass range $m_\chi \lesssim$ 2 GeV$/c^2$. For decays into photons, HERA improves on other searches in the low mass range for $m_\chi\lesssim$ few MeV$/c^2$.

In the light of these very good prospects, it would be necessary to refine this study making use of a more advanced statistical analysis such as e.g., Bayesian inference. The latter is much more time consuming/computationally expensive but would give rise to more realistic estimates of the prospect for constraining DM decays. We will also vary the complexity of the astrophysical models, comparing the Bayesian evidences (e.g., \cite{Qin2020MNRAS.495..123Q}). In addition, DM could also heat the IGM through annihilations. In this case, one should  take special care in the treatment of the late time boost arising for structure formation.  This induces yet another modeling uncertainty, but is worth exploring given the sensitivity of upcoming 21-cm measurements to IGM heating during the CD.

\acknowledgments

We thank S. Junius for collaboration at early stages of this work as well as T.~Slatyer and Q.~Decant for useful discussions on 21cm cosmology and DM imprint. GF acknowledges support of the ARC program of the Federation Wallonie-Bruxelles and of the Excellence of Science (EoS) project No. 30820817 - be.h “The H boson gateway to physics beyond the Standard Model”. LLH is supported
by the Fonds de la Recherche Scientifique F.R.S.-FNRS through a research associate position and acknowledges support of the  FNRS research grant number F.4520.19, the ARC program of the Federation Wallonie-Bruxelles and  the IISN convention No. 4.4503.15.
Computational resources have been provided by the Consortium des Equipements de Calcul Intensif (CECI), funded by the Fonds de la Recherche Scientifique de Belgique (F.R.S.-FNRS) under Grant No. 2.5020.11 and by the Walloon Region of Belgium. YQ is supported by the Australian Research Council Centre of Excellence for All Sky Astrophysics in 3 Dimensions (ASTRO 3D), through project \#CE170100013. Part of this work was performed on the OzSTAR and Gadi national computational facilities in Australia.

\appendix

\section{ Extra information on IGM heating sources}
\label{app:MCGs_LX_DM}

Here we provide further details on X-ray energy injection in \refapp{app:Xrays} and we discuss the impact of MCGs   and the DM heating when considering both ACGs and MCGs in \refapp{app:MCGs}.

\subsection{X-ray heating contributions}
\label{app:Xrays}

The  X-ray emissivity $\epsilon_X$ introduced in \refequ{eq:e_X} contributes to  secondary ionisations, heating and the Lyman alpha background appearing in eqs.~(\ref{eq:xe}) and~(\ref{eq:TK}). We first introduce  the total integrated radiation intensity
\begin{equation}
I_{\rm X} =\frac{\left(1+z\right)^3}{\rm 4{\rm \pi}} \int_{z}^{\infty} {\rm d}z^\prime\frac{{c \rm d}t}{{\rm d}z^\prime} {\epsilon}_{\rm X}\exp(-\tau_{\rm X}),
\label{eq:J_LW}
\end{equation}
where $\tau_{\rm X}$ stands for the X-rays optical depth in the IGM (see more in Ref~\citep{Mesinger:2010ne}). The latter depends on the specific X-ray luminosity per SFR whose amplitude and softness/hardness is set by the parameters $L^{\rm I/II}_X$ and $E_0$, see \refequ{eq: LX}. The ionization and heating rates from X-ray photons (i.e., $\Lambda_{\rm ion}^{\rm X}$ in~\refequ{eq:xe} and $\epsilon_{\rm heat}^{\rm X}$ in~\refequ{eq:TK}) as well as their contribution to the Lyman-$\alpha$ radiation (i.e., $J_{\alpha}^X$ in equation \ref{eq:Jalpha}) follows
\begin{equation}
    \Lambda_{\rm ion}^{\rm X} = \int_{E_0}^{\infty} {\rm d}E \frac{4{\rm \pi} I_X}{E} \sum_{j}  x_{j} \sigma_{j}  \mathfrak{f}_{j} \left[{(E{-}E_{\rm th}^{j})}\sum_{k} \frac{f_{\rm ion}^{k}}{E_{\rm th}^{k}} + 1 \right],
\end{equation}
\begin{equation}
    \epsilon_{\rm heat}^{X} = \int_{E_0}^{\infty} {\rm d}E \frac{4{\rm \pi} I_X}{E} \sum_{j}  x_{j} \sigma_{j} \mathfrak{f}_{j}{(E{-}E_{\rm th}^{j})} f_{\rm heat},
\end{equation}
and
\begin{equation}
J_{\alpha}^{\rm X} =  \frac{c n_{\rm b}}{4{\rm {\rm \pi}}H\left(z\right)\nu_\alpha} \int_{E_0}^{\infty} \mathrm{d}E \frac{4{\rm \pi} I_X}{E} \sum_{j}  x_{j} \sigma_{j} \mathfrak{f}_{j}{(E{-}E_\mathrm{th}^{j})} \frac{f_{\alpha}}{h\nu_\alpha}.
\end{equation}
Here, $\mathfrak{f}_{j}$ refers to the number fraction of each species, $j$, in HI, HeI,and HeII, with $\sigma_{j}$ and $E_\mathrm{th}^{j}$ being their corresponding cross-section and energy for ionization; $f_\mathrm{heat}$, $f_\mathrm{ion}^{k}$ and $f_{\alpha}$ represent the fraction of the electron energy after ionization, $E{-}E_\mathrm{th}^{j}$, that contributes to heating, secondary ionization or emitting Lyman-$\alpha$ photons (${\nu_\alpha}\equiv 2.47\times10^{15}\mathrm{Hz}$) of each species \citep{Furlanetto2010MNRAS.404.1869F}; and $x^{j} \equiv 1{-} x_e$ for HI and HeI or $x_e$ for HeII represents the secondary ionization fractions (See more in \cite{Qin2020MNRAS.495..123Q}).

\subsection{MCGs vs DM decays imprint on 21cm signal}
\label{app:MCGs}

\begin{figure}[t!]
\centering
\includegraphics[width=0.46\linewidth]{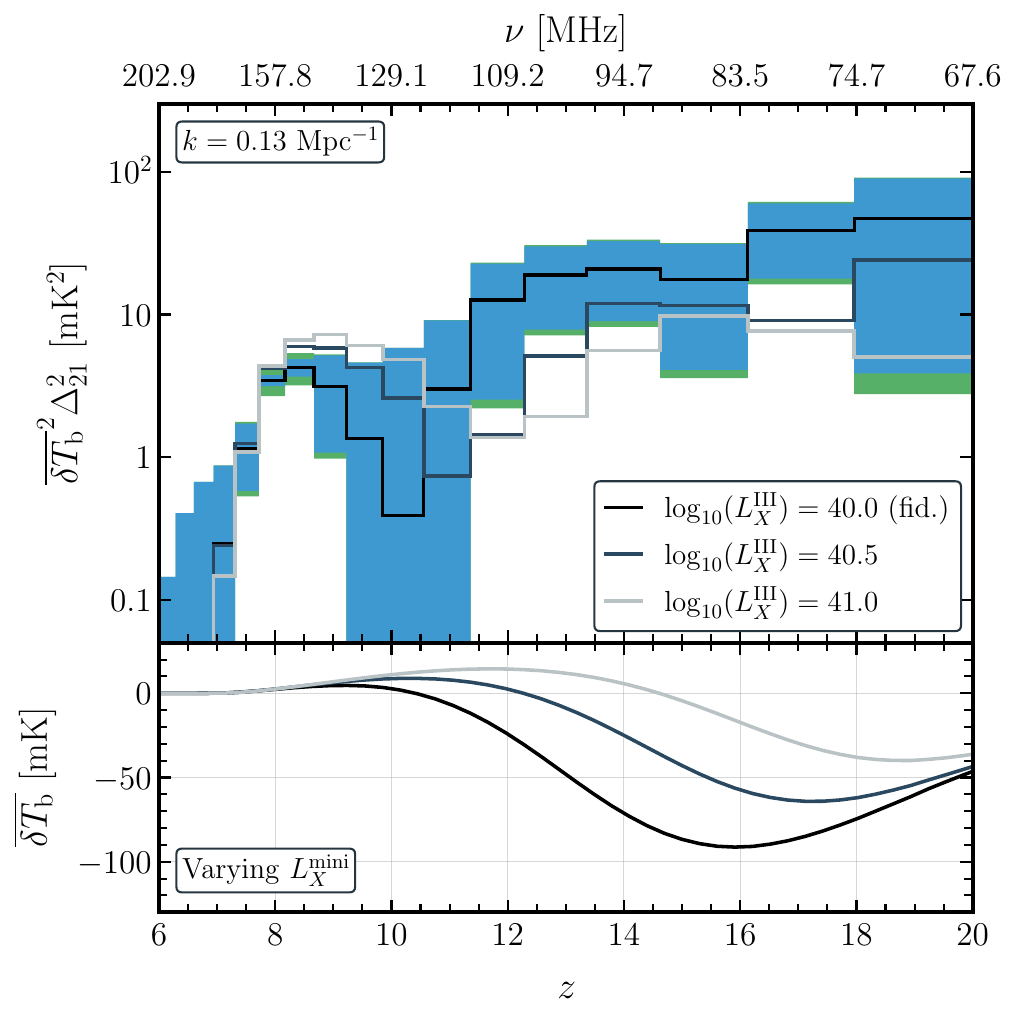}
\includegraphics[width=0.46\linewidth]{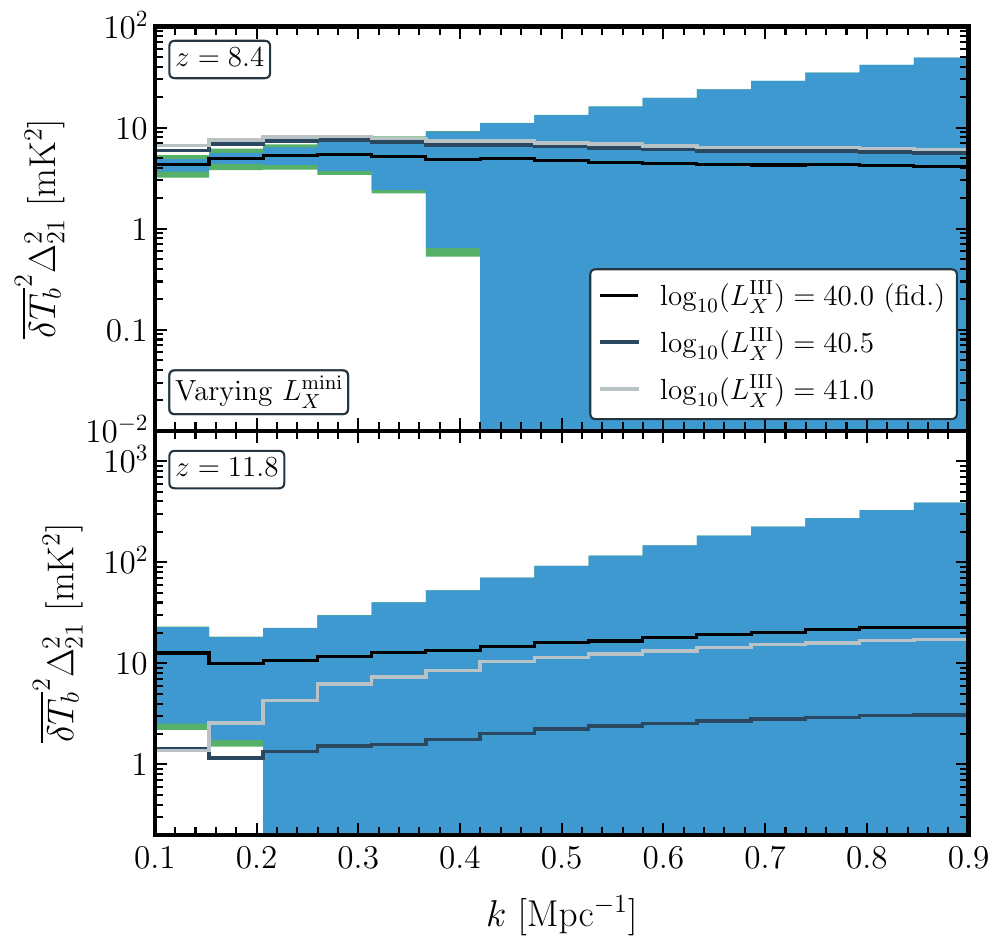}
\caption{Similar as \reffig{fig:global-PS-LX} including MCGs and varying   $L_X^{\rm III}$. We show the 1$\,\sigma$ error (instead of $2\, \sigma$) for a more readable figure.}
\label{fig:global-PS-LXmini}
\end{figure}
\begin{figure}
\centering
\includegraphics[width=0.46\linewidth]{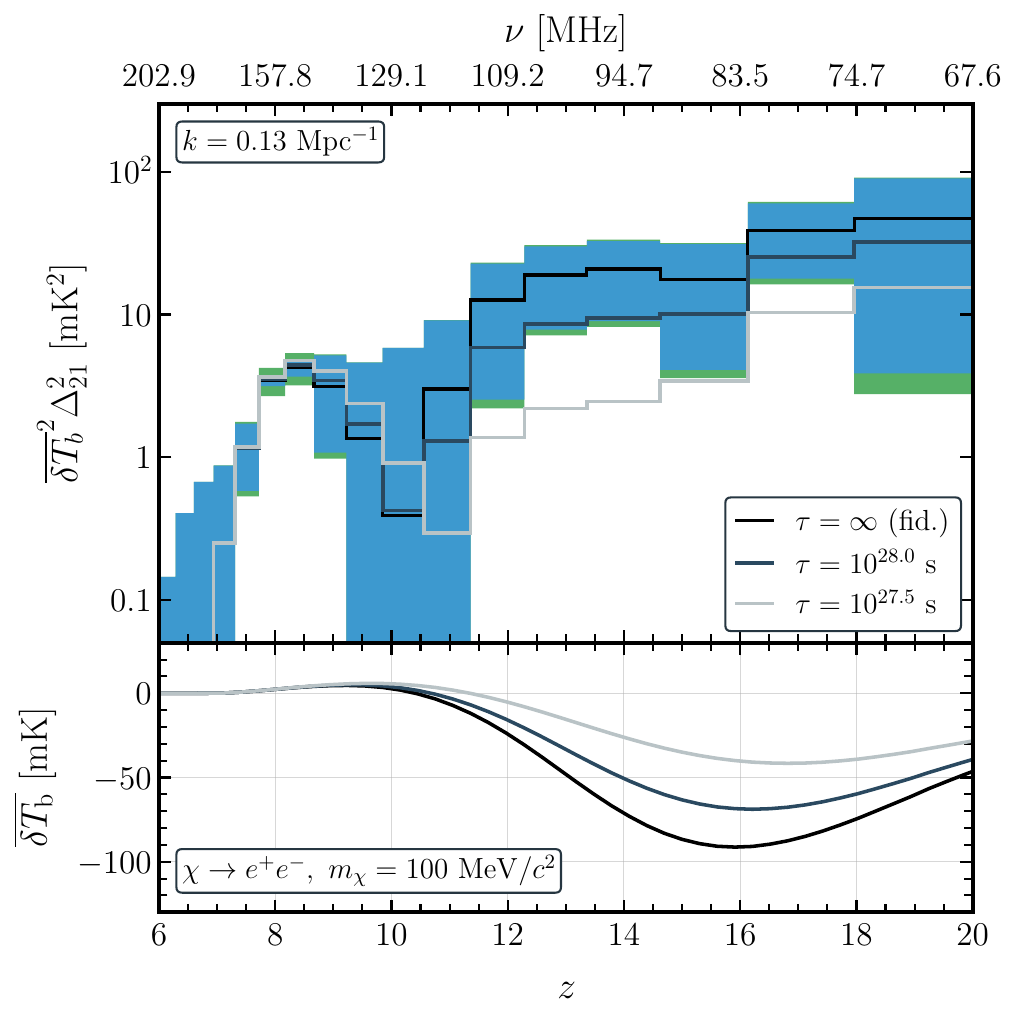}
\includegraphics[width=0.46\linewidth]{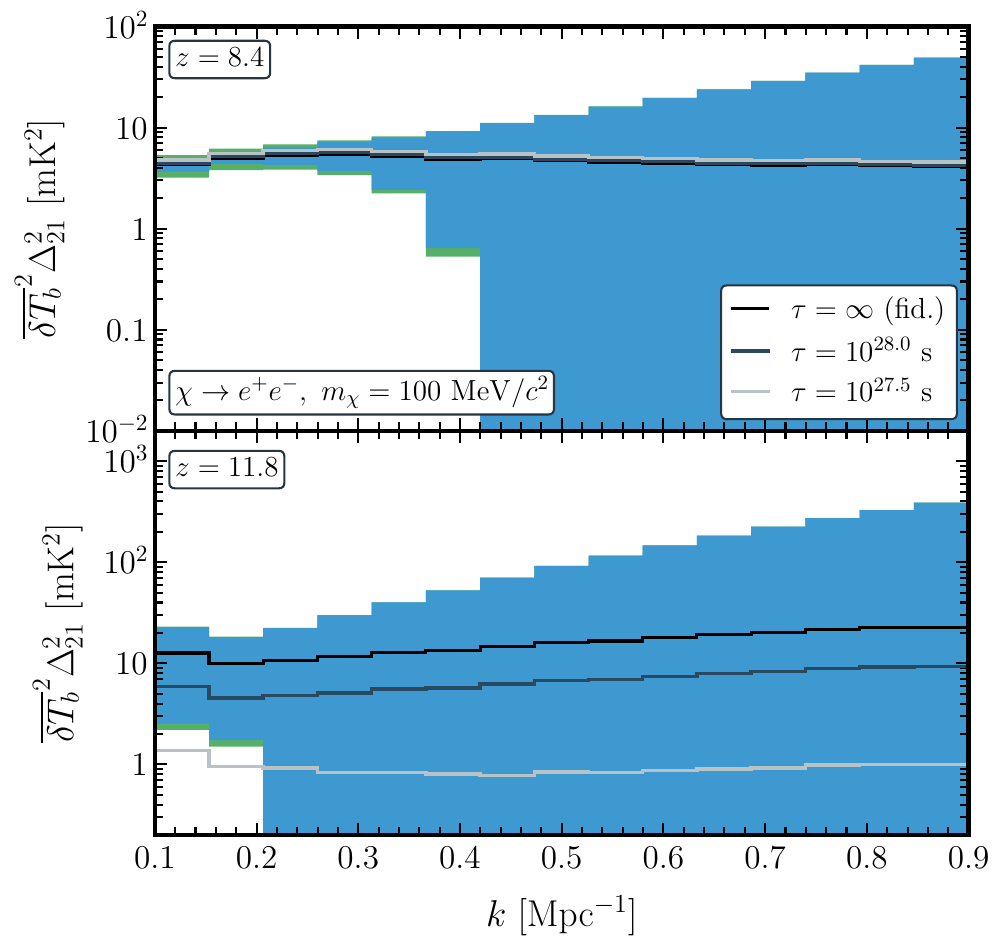}
\caption{Same as \reffig{fig:global-PS-LX-DM} including MCGs and showing the 1$\,\sigma$ error (instead of $2\, \sigma$ for a more readable figure).}
\label{fig:global-PS-DMmini}
\end{figure}

\begin{figure}[t!]
    \centering
    \includegraphics[width=0.95\linewidth]{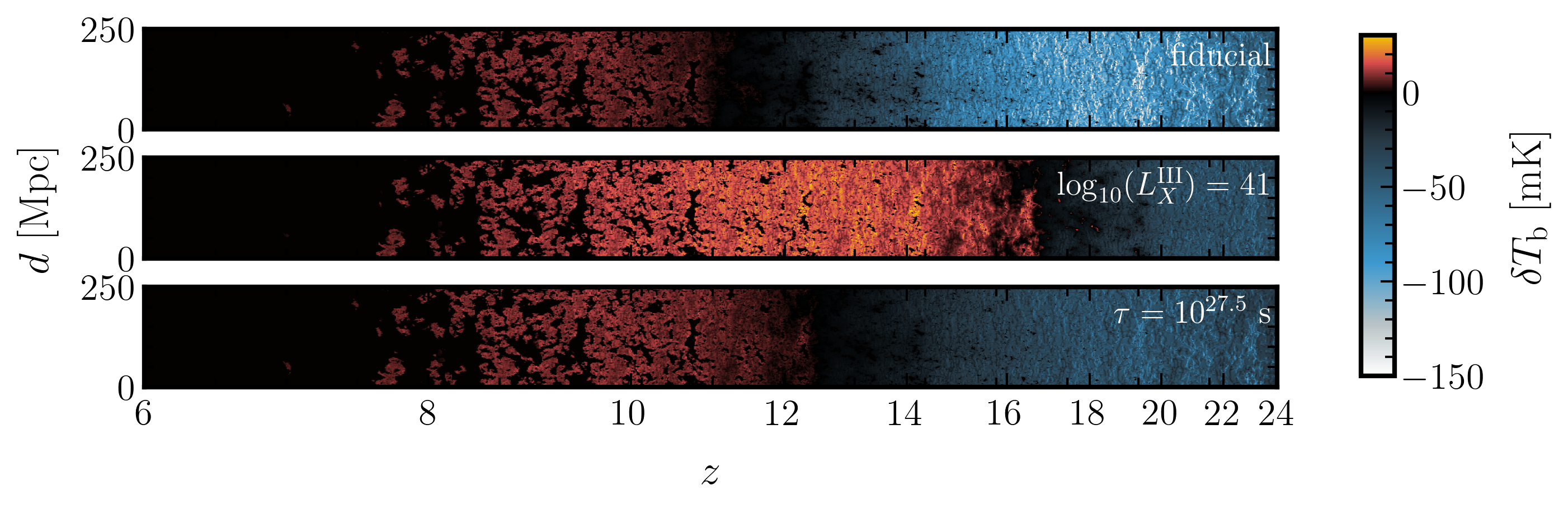}
    \caption{Time slices of the differential brightness temperature in our $(250~{\rm Mpc})^3$ large simulation box . We show results for the ACGs + MCGs fiducial model -- with $\log_{10}(L_X^{\rm III}) = 40$ and $\tau = \infty$ -- (upper panel), $\log_{10}(L_X^{\rm III}) = 41$ (middle panel), and for a 100 MeV$/c^2$ DM with $\tau = 10^{27.5}~{\rm s}$ (bottom panel).}
    \label{fig:lightcones_minihalos}
\end{figure}

As shown in \reffig{fig:heating_rates}, the X-ray heating rate from \popIII{}-dominated MCGs decreases more smoothly in redshift than that from ACGs. This is  similar to the case of the DM heating rate that is roughly constant. It is thus expected that the heating parameters associated to MCGs, including $L_X^{\rm III}$, shall be more strongly  degenerate with $\Gamma$, which normalises  the DM decay rate, than the ones associated to ACGs.  Figures~\ref{fig:global-PS-LXmini} and ~\ref{fig:global-PS-DMmini}, illustrating the dependence of the 21cm signal on $L_X^{\rm III}$ and $\Gamma$ in an ACGs+MCGs scenario, can  be compared to figures \ref{fig:global-PS-LX} and \ref{fig:global-PS-LX-DM} for  ACGs only. Including MCGs, we see that the fiducial scenario (black curve) present  a shallower and earlier absorption  in the global signal. Furthermore, the power spectrum is suppressed  at larger redshifts (between redshifts 12 and 16) and the EoR peak is wider (between redshifts 6 and 10). Figure~\ref{fig:global-PS-LXmini} illustrates the effect of varying  $L_X^{\rm III}$, instead of  $L_X^{\rm II}$ in  \reffig{fig:global-PS-LX}, on the 21cm signal while~\reffig{fig:global-PS-DMmini} shows variations in the signal due to the DM life time. In the MCGs+ACGs case, it is more difficult to disentangle DM imprint from astrophysics.  Indeed,   the features in the power spectrum that can help discriminate the DM heating from first galaxies X-ray heating are pushed to larger redshifts (where the instrument is less sensitive) and  are more difficult to detect due to a smaller signal-to-noise ratio. In figures~\ref{fig:global-PS-LXmini} and \ref{fig:global-PS-DMmini}, we show the noise at 1$\sigma$ level with the blue (experimental noise only) and green (total noise) bars -- while it was shown at 2$\sigma$ in figures \ref{fig:global-PS-LX} and \ref{fig:global-PS-LX-DM}. Nonetheless, we observe that modifying the fiducial values of $L_X^{\rm III}$ and $\tau$ to $10^{40.5}$ and $10^{27.5}$ s are already sufficient to shift the signal outside the $1\,\sigma$ uncertainty area for some of the probed redshifts and $k$ values. We also  provide time slices of the differential brightness temperature in the MCGs+ACGs scenario in \reffig{fig:lightcones_minihalos} that can be compared to \reffig{fig:lightcones}. 

\section{Fisher matrix analysis}
\label{app:FisherGlobal}

In this appendix we give more details on the mathematics behind the Fisher analysis, the binning choice we choose and the treatment of the decay parameter. Furthermore, we provide full triangle plots for all parameters considered in the analysis.

\subsection{General method}
\label{app:Fisher}

\begin{table}[t!]
    \centering
    \begin{tabular}[t]{ c | c  c | c }
        $n$ & $\hat z_{n-1}~(\frac{\hat \nu_{n-1}}{\rm MHz})$ & $\hat z_n~(\frac{\hat \nu_{n}}{\rm MHz})$ &  $z_n~(\frac{\nu_{n}}{\rm MHz})$ \\
        \hline
        1 & 6.00~(202.9) & 6.29~(194.9) & 6.14~(198.9) \\
        2 & 6.29~(194.9) & 6.60~(186.9) & 6.44~(190.9) \\
        3 & 6.60~(186.9) & 6.94~(178.9) & 6.77~(182.9) \\
        4 & 6.94~(178.9) & 7.31~(170.9) & 7.12~(174.9) \\
        5 & 7.31~(170.9) & 7.72~(162.9) & 7.51~(166.9) \\
        6 & 7.72~(162.9) & 8.17~(154.9) & 7.94~(158.9) \\
        7 & 8.17~(154.9) & 8.67~(146.9) & 8.41~(150.9) \\
        8 & 8.67~(146.9) & 9.22~(138.9) & 8.94~(142.9) \\
        9 & 9.22~(138.9) & 9.85~(130.9) & 9.53~(134.9) \\
        10 & 9.85~(130.9) & 10.56~(122.9) & 10.19~(126.9) \\
        11 & 10.56~(122.9) & 11.36~(114.9) & 10.94~(118.9) \\
        12 & 11.36~(114.9) & 12.29~(106.9) & 11.81~(110.9) \\
        13 & 12.29~(106.9) & 13.36~(98.9) & 12.80~(102.9) \\
        14 & 13.36~(98.9) & 14.62~(90.9) & 13.97~(94.9) \\
        15 & 14.62~(90.9) & 16.13~(82.9) & 15.34~(86.9) \\
        16 & 16.13~(82.9) & 17.96~(74.9) & 17.00~(78.9) \\
        17 & 17.96~(74.9) & 20.23~(66.9) & 19.03~(70.9) \\
    \end{tabular}
    \hspace*{10pt}
    \begin{tabular}[t]{ c | c c | c }
        $n$ & $\frac{\hat k_{n-1}}{{\rm Mpc^{-1}}}$ & $\frac{\hat k_n}{{\rm Mpc^{-1}}}$ & $\frac{k_n}{{\rm Mpc^{-1}}}$ \\
        \hline
        1 & 0.10 & 0.15 & 0.13 \\
        2 & 0.15 & 0.21 & 0.18 \\
        3 & 0.21 & 0.26 & 0.23 \\
        4 & 0.26 & 0.31 & 0.29 \\
        5 & 0.31 & 0.37 & 0.34 \\
        6 & 0.37 & 0.42 & 0.39 \\
        7 & 0.42 & 0.47 & 0.45 \\
        8 & 0.47 & 0.53 & 0.50 \\
        9 & 0.53 & 0.58 & 0.55 \\
        10 & 0.58 & 0.63 & 0.61 \\
        11 & 0.63 & 0.69 & 0.66 \\
        12 & 0.69 & 0.74 & 0.71 \\
        13 & 0.74 & 0.79 & 0.77 \\
        14 & 0.79 & 0.85 & 0.82 \\
        15 & 0.85 & 0.90 & 0.87 \\
        16 & 0.90 & 0.95 & 0.93 \\
    \end{tabular}
        \caption{{\bf Left panel}. Redshift bin edges $\hat z_{\rm n}$ and associated centers $z_n$. In parenthesis we put the associated frequencies of the redshifted 21cm line. {\bf Right panel}. Mode bin edges $\hat k_{\rm n}$ and associated centers $k_n$.}
    \label{tab:zkbins}
\end{table}

Here we detail the formalism of the Fisher matrix inference method. Our analysis is inspired and similar to that introduced in ref.~\cite{Mason:2022obt}. For the following discussion we assume that the power spectrum is binned into a grid of $N_k$ $k$-modes over $N_z$ redshifts. We define the vector of length $N \equiv N_k \times N_z$, 
\begin{equation}
   \boldsymbol X \equiv (\overline{\delta T_b}^2(z_1)\Delta^2_{21}(k_1, z_1), \overline{\delta T_b}^2(z_1)\Delta^2_{21}(k_1, z_1), \dots, \overline{\delta T_b}^2(z_{N_z})\Delta^2_{21}(k_{N_k}, z_{N_z})) 
\end{equation}
capturing all the information that can experimentally be obtained about the 21cm power-spectrum. From measured $\boldsymbol X$, one could reconstruct the posterior distribution on the parameters using an inference algorithm. However, because we aim at forecasting sensitivities of forthcoming experimental runs, one can only rely on the theoretical predictions obtained from \cmFAST{} and that depend on the choice of model to describe the Universe. A Fisher analysis is thus adequate to obtain reliable estimates in a relatively short time (without requiring extensive computational resources). 

We assume that $\boldsymbol X$ is randomly distributed according to the likelihood $\mathcal{L} : (\boldsymbol X, \boldsymbol \theta) \mapsto \mathcal{L}(\boldsymbol X\, | \, \boldsymbol \theta) $ with $\boldsymbol \theta$ a vector of parameters. The Fisher information matrix $F$ associated to $\mathcal{L}$ and evaluated for the parameters $\boldsymbol \theta$ is given by
\begin{equation}
    F_{ij}  = - \left< \frac{\partial^2 \ln \mathcal{L}}{\partial \theta_i \partial \theta_j}\right> \equiv -\mathbb{E}_{\boldsymbol X}\left[ \left. \frac{\partial^2}{\partial \theta_i \partial \theta_j} \ln \mathcal{L}(\boldsymbol X \, | \, \boldsymbol \theta)\, \right| \, \boldsymbol \theta \right]
    \label{eq:Fisher_information}
\end{equation}
where $\mathbb{E}_{\boldsymbol X}$ is the expectation value associated to $\mathcal{L}$. The Cramér-Rao limit \cite{frechet1943extension, darmois1945limites, aitken1942xv} 
ensures that the covariance matrix associated to the posterior distribution on the set of parameters $\boldsymbol \theta$ satisfies the inequality
\begin{equation}
   (\mathcal{C}_{\theta})_{ij} \ge (F^{-1})_{ij} \quad \forall (i,j) \, .
\end{equation}
We assume the optimistic case and define the covariance of the posterior as the inverse of the Fisher matrix. We consider that $\boldsymbol X$ follows a multivariate Gaussian distribution with covariance matrix  $\mathcal{C}_X$ (independent of $\boldsymbol \theta$)\footnote{If the covariance matrix were dependent on the parameters $\boldsymbol \theta$, the Fisher matrix element given in \refequ{eq:FisherElement} would have additional terms depending on $\partial \mathcal{C}_X / \partial \theta_i$.} and mean $\boldsymbol \mu_{\boldsymbol X}(\boldsymbol \theta)$: $\boldsymbol X \sim \mathcal{N}(\boldsymbol \mu_{\boldsymbol X}(\boldsymbol \theta), \mathcal{C}_X)$. Thus, the likelihood and the Fisher matrix elements take the form,
\begin{equation}
\begin{split}
      \mathcal{L}(\boldsymbol x \, | \, \boldsymbol \theta) = & \frac{e^{-\frac{1}{2} \left[{\boldsymbol x}- {\boldsymbol \mu}_{\boldsymbol X}(\boldsymbol \theta)\right]^{\rm T} \mathcal{C}_X^{-1}  \left[\boldsymbol x- \boldsymbol\mu_{\boldsymbol X}(\boldsymbol \theta)\right] } }{\sqrt{(2\pi)^{N} |{\rm det} \, \mathcal{C}_X|}}  \quad  {\rm and} \quad F_{ij} = \frac{\partial \boldsymbol \mu_{\boldsymbol X}^{\rm T}(\boldsymbol \theta)}{\partial \theta_i} \mathcal{C}_X^{-1} \frac{\partial \boldsymbol \mu_{\boldsymbol X}(\boldsymbol \theta)}{\partial \theta_j} \, ,
      \label{eq:FisherElement}
\end{split}
\end{equation}
where T indicates vector transposition. Therefore, we can estimate the covariance matrix of the parameters $\boldsymbol \theta$ from a matrix-sum on the derivative of the mean value of the data. Having no real data set at our disposal, we can only assume that the true Universe is similar to that simulated in \cmFAST{} for a chosen fiducial model $\boldsymbol \theta = \boldsymbol \theta_{\rm fid}$. Hence, $\boldsymbol \mu_{\boldsymbol X}$ is directly given by the binned output obtained from that fiducial model and one computes derivatives by slightly varying the parameters around the fiducial values.

\subsection{Details on the binning choice}
\label{app:binning}

Notwithstanding that \cmCAST{} can work for different binning choice, by default, we bin the mock data according to the frequency bandwidth $B$ of the instrument. The redshift bin width is straightforwardly related to the frequency bandwidth through the redshifted 21cm line frequency $\nu(z) = \nu_{21}/(1+z)$. We write the minimum redshift considered for the analysis as $z_{\rm min} = \hat{z}_0$ and define frequency bins with edges $\{\hat{\nu}_n\}_{n \ge 0}$ and centers $\{\nu_n\}_{n \ge 1}$ by
\begin{equation}
    \begin{cases}
        \hat{\nu}_n = \nu(\hat z_0) - nB \\
        \nu_n = \nu(\hat z_0)  - \left(n-\frac{1}{2}\right)B \, .
    \end{cases}
\end{equation}
Subsequently, $z$-bins are defined with edges $\hat z_n = \nu_{21}/\hat \nu_n -1$ and \emph{centers} $z_n = \nu_{21}/\nu_n -1$. The sequence runs up to $n=N_z$, until $\hat z_{N_z+1}$ becomes greater than a fixed threshold $z_{\rm max}$.  A mode $k$ is given by the quadratic sum of its two components, $k_{\parallel}$ and $\boldsymbol k_\perp$ respectively parallel and perpendicular to the line of sight. The minimal accessible value of $|\boldsymbol k_\perp|$ is fixed by the minimal baseline distance between the antennas. However, that of $k_\parallel$ happens to be orders of magnitude larger and is fixed by the bandwidth \cite{HERA:2021bsv}. One can associate a frequency range $\Delta \nu$ to a range of the parallel projection of modes using
\begin{equation}
    \Delta k_\parallel(z, \Delta \nu) \equiv 2\pi \frac{\nu_{21}}{\Delta \nu} \frac{H(z)}{(1+z)^2} \, ,
\end{equation}
 decreasing function of the redshift. Thus, we define the $k$-bin width from the maximal possible value of $\Delta k_\parallel$ at $\hat z_0$ for $\Delta \nu = B$. On a range $[k_{\rm min}, k_{\rm max}]$, the bin edges are
\begin{equation}
\hat k_n \equiv {\rm max} \left\{ k_{\rm min}, \Delta k_{\parallel}(\hat z_0, B)  \right\} + n \Delta k_{\parallel}(\hat z_0, B) \, ,   
\end{equation}
up to  $n = N_k$ where $\hat k_n$ becomes larger than $k_{\rm max}$. The instrument being limited by its spectral resolution $\delta \nu$, $k_{\rm max}$ can not be arbitrary and should at least satisfy $k_{\rm max} < \Delta k_\parallel(z_{\rm max}, \delta \nu)/2$.  \\

\begin{figure}[t!]
    \centering
    \includegraphics[width=0.99\linewidth]{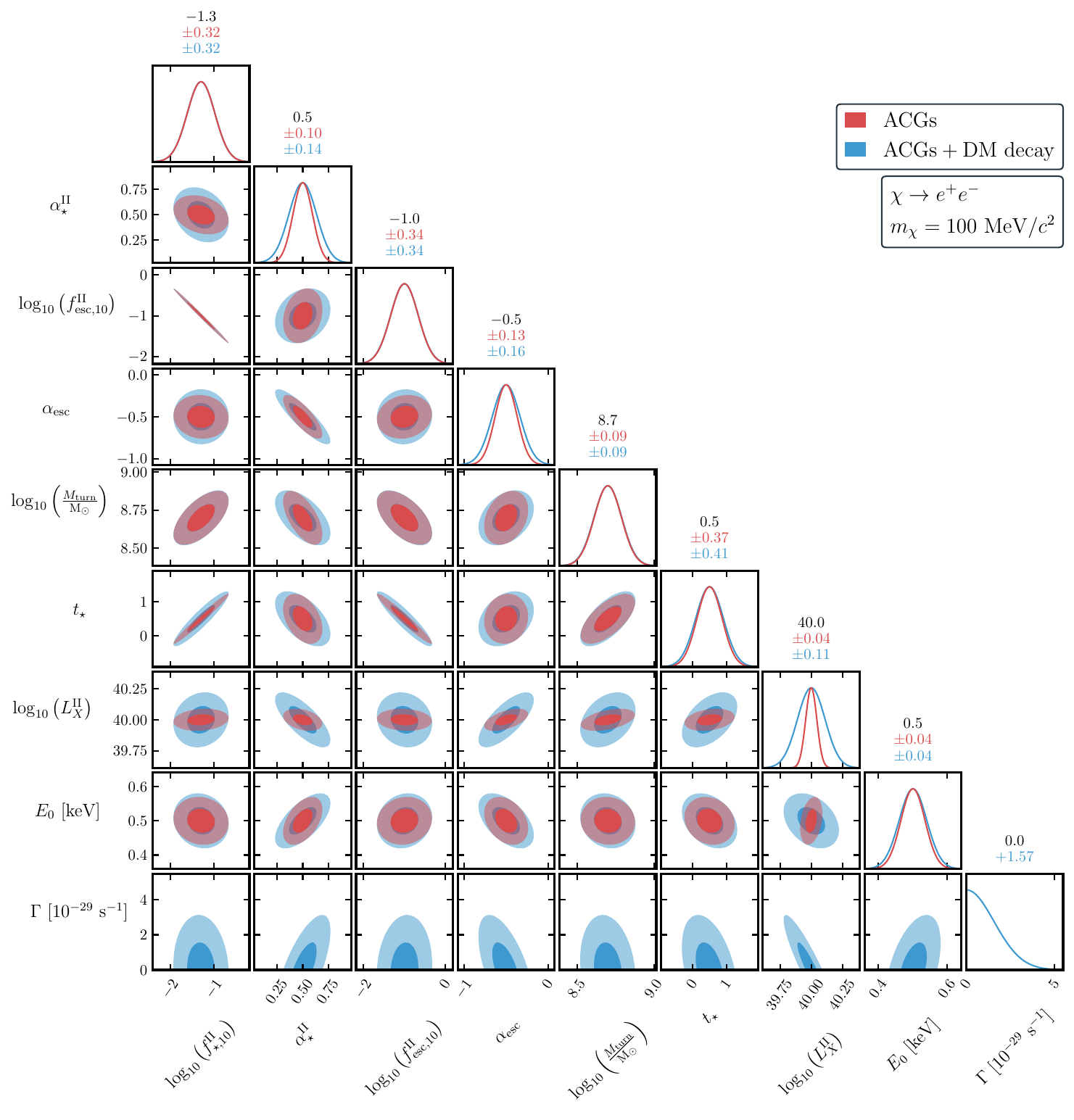}
    \caption{{\small Triangle plot result of the Fisher analysis with the contribution of a single polulation of \popII{} dominated ACGs. In blue we consider a 100 MeV$/c^2$ DM decaying into electron positron pairs. For comparison, in red we show the result of the Fisher forecast without DM decay. We can conclude that DM decay mostly impacts on the reconstruction of $L_X^{\rm II}$.} }
    \label{fig:Triangle}
\end{figure}

We put in \reftab{tab:zkbins} the fixed grid in redshifts and modes that we use to bin the output of \cmFAST{} and perform the Fisher matrix analysis. Note that it appears we could also have simply defined the center of the redshift bins $z_n$ as
\begin{equation}
    \tilde z_n = \frac{\hat z_n+ \hat z_{n-1}}{2} \quad {\rm since} \quad
    \frac{ \tilde z_n }{z_n} = \frac{(\hat \nu_n + 1/2)^2}{\hat \nu_n(\hat \nu_n +1)}
\end{equation}
goes to 1 when $\hat \nu_n \gg 1$. Indeed, in practice $\hat \nu_n > 65$ for the redshift range we consider, thus the two definitions are completely equivalent.

\subsection{Treatment of dark matter decay in the Fisher forecast}
\label{app:DMdecay}

\begin{figure}[t!]
    \centering
    \includegraphics[width=0.99\linewidth]{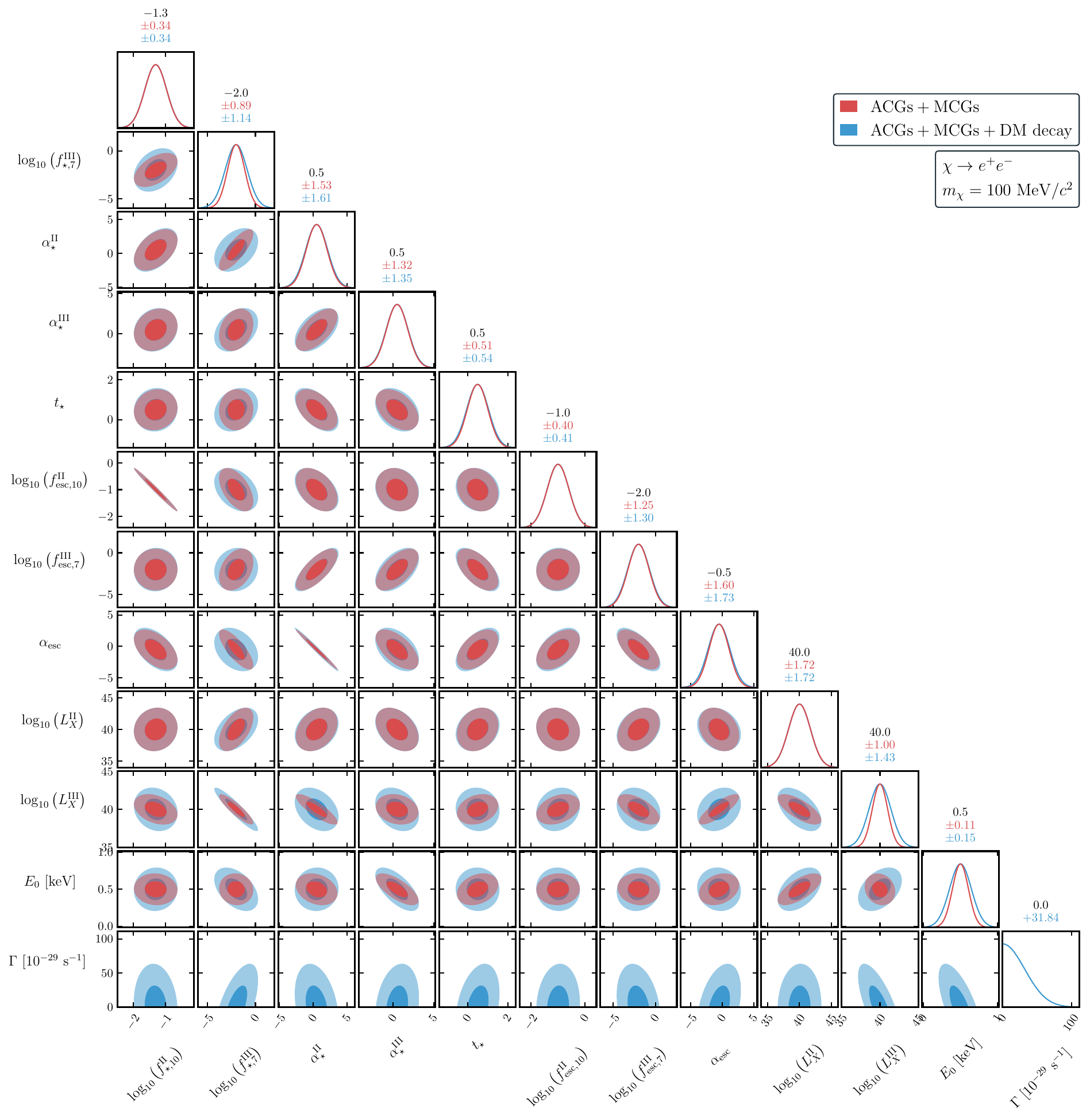}
    \caption{{\small Same as \reffig{fig:Triangle} but adding the contribution of \popIII{}-dominated MCGs. Here, DM decay impacts more strongly the reconstruction of $L_{X}^{\rm III}$ than the one of $L_{X}^{\rm II}$.}}
    \label{fig:TrianglePlotPOPIII}
\end{figure}

The constraint on the DM lifetime $\tau$, or equivalently its decay rate $\Gamma$, is derived assuming a fiducial model with no exotic energy injection. We thus have a fiducial value of the decay rate set to $\Gamma_{\rm fid} = 0~{\rm s}^{-1}$. That is to say, we want to quantify by how much $\Gamma$ can be larger than zero before it cannot be neglected in the parameter reconstruction. This gives rise to technical difficulties in the numerical evaluation of the derivative of the power spectrum with respect to the decay rate -- see \refequ{eq:FisherMatrix}. Firstly, because the fiducial value for $\Gamma$ equal zero,\footnote{Notice that  $\tau$ is even more difficult to handle in a numerical analysis than $\Gamma$ as a DM candidate that is not decaying has infinite lifetime, i.e., its fiducial value would be $\tau_{\rm fid}=\infty$.} one cannot consider its logarithm even-though $\Gamma$ has to vary on many orders of magnitude. For the that reason, varying the parameters by a few percent to compute the derivative also does not make sense here. Secondly, it is not possible to evaluate the power spectrum for $\Gamma < 0$. As a result, we use a one side derivative scheme (instead of a centered scheme for the other parameters),
\begin{equation}
    \left. \frac{\partial f}{\partial \Gamma} \right|_{\Gamma = 0} = \frac{f(\epsilon_\Gamma) - f(0)}{\epsilon_\Gamma} \, ,
\label{eq:dfdGam}
\end{equation}
to evaluate the derivative of a function $f$ with respect to $\Gamma$,  with $\epsilon_\Gamma$, a small parameter. In practice here we have $f=\overline{\delta T_b}^2\Delta_{21}^2$.

The choice of `small value' for $\epsilon_{\Gamma}$ used in the numerical analysis is quite critical as $\Gamma$ vary over orders of magnitude and one needs $\partial f/\partial \Gamma$ to be independent of the choice of $\epsilon_{\Gamma}$.
Indeed, for choices of $\epsilon_{\Gamma}$  too small, $f(\epsilon_\Gamma)$ is expected to be essentially  identical to $f(0)$ up to the numerical noise and one has then $\partial f / \partial \Gamma \propto 1/\epsilon_\Gamma$. On the other hand, when $\epsilon_{\Gamma}$ is  too large, the expression \eqref{eq:dfdGam}  is no longer valid for a derivative. In order to make a sensible choice, we have computed the marginalized error on $\Gamma$, $\sigma_\Gamma = \sqrt{(F^{-1})_{i_\Gamma i_\Gamma}}$, for various values of $\epsilon_\Gamma$. We have then selected a value of $\epsilon_ \Gamma$ at which  $\sigma_\Gamma$ is independent $\epsilon_\Gamma$, i.e., for which ${\rm d} \sigma_\Gamma / {\rm d} \epsilon_\Gamma \sim 0$. In the analysis performed in this work, we find that the correct choice for $\epsilon_\Gamma$ is a few percent of the error $\sigma_\Gamma$, {\emph i.e}, $  \epsilon_\Gamma=10^{-27} - 10^{-29}~{\rm s^{-1}}$. 

\begin{figure}[t!]
    \centering
    \includegraphics[width=0.99\linewidth]{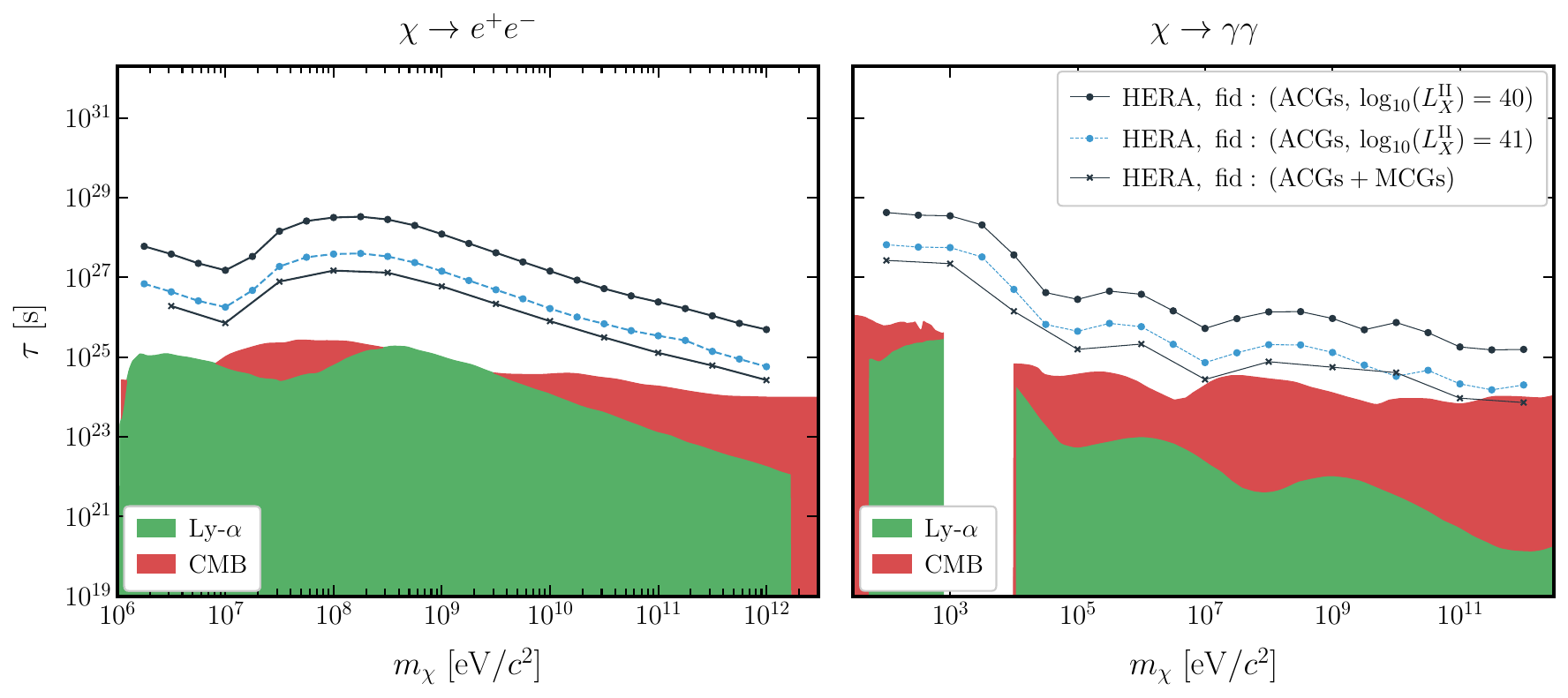}
    \caption{{\small Constraints on the dark matter lifetime (at 95\% level) for decay into an electron/positron pair (left panel) and photons (right panel). We superpose the forecasts for the HERA telescope assuming \popII{}-dominated ACGs only and $\log_{10}(L_X^{\rm II}) = 41$ (dashed blue line with round markers), \popII{}-dominated ACGs only and $\log_{10}(L_X^{\rm II}) = 40$ (solid dark line with round markers), or \popII{}-dominated ACGs  + \popIII{}-dominated MCGs with $\log_{10}(L_X^{\rm II,III}) = 40$  (solid dark line with crosses) with existing cosmological constraints. We refer to \reffig{fig:decaybound} for more details.}}
    \label{fig:constraints_LX41}
\end{figure}

\subsection{Full triangle plots}
\label{app:Fisher_triangle_plots}

The full triangle plots resulting from our Fisher Matrix forecasts are shown in~\reffig{fig:Triangle}, completing the left plot of \reffig{fig:Triangle_Zoom} for the analysis involving \popII{}-dominated ACGs. Furthermore, \reffig{fig:TrianglePlotPOPIII}, completes the right plot of \reffig{fig:Triangle} for the analysis involving both \popII{}-dominated ACGs and \popIII{}-dominated MCGs.

\section{Impact of the X-ray normalisation on the forecast}
\label{app:impact_LX_forecast}

In this section we show the results of our Fisher analysis in the ACGs-only scenario but considering $\log_{10}(L_X^{\rm II}) = 41$ (instead of 40) as the fiducial value. In \reffig{fig:constraints_LX41} the 95\% CL bound for the new fiducial value of $L_X^{\rm II}$ is shown (dashed blue line with round markers) together with the other cosmological constraints from CMB and Lyman-$\alpha$. For comparison, the bounds obtained in the ACGs-only scenario with $\log_{10}(L_X^{\rm II}) = 40$ (solid dark line with round markers) and in the ACGs+MCGs scenario with $\log_{10}(L_X^{\rm II/III}) = 40$ (solid dark line with crosses) are also plotted. With a larger X-ray injection the contribution of dark matter decay is harder to disentangle from the astrophysical sources (see \reffig{fig:heating_rates}). Consequently, there is a drop of almost an order of magnitudes between the fiducial $\log_{10}(L_X^{\rm II}) = 40$ and the fiducial $\log_{10}(L_X^{\rm II}) = 41$. Nonetheless, the constraints with ACGs and $\log_{10}(L_X^{\rm II}) = 41$ remains slightly stronger than considering MCGs and than the CMB and Lyman-$\alpha$ bounds.

\addcontentsline{toc}{section}{Bibliography} 
\bibliographystyle{JHEP}
\bibliography{main} 

\end{document}